\documentclass[a4paper,12pt]{article}
\pdfoutput=1
\usepackage{epsfig}
\usepackage{amssymb}
\usepackage{amsfonts}
\usepackage{amsmath}
\usepackage{euscript}
\usepackage{verbatim}
\usepackage{latexsym}
\usepackage{graphicx}
\usepackage{caption}
\usepackage{float}
\usepackage{subcaption}
\usepackage{bbm}

\usepackage{listings}
\newcommand{\code}[1]{\texttt{#1}}

\usepackage{booktabs}

\newif\ifdtup

\catcode`\@=11

\@addtoreset{equation}{section}

\def\@normalsize{\@setsize\normalsize{15pt}\xiipt\@xiipt
\abovedisplayskip 14pt plus3pt minus3pt%
\belowdisplayskip \abovedisplayskip
\abovedisplayshortskip \z@ plus3pt%
\belowdisplayshortskip 7pt plus3.5pt minus0pt}

\def\small{\@setsize\small{13.6pt}\xipt\@xipt
\abovedisplayskip 13pt plus3pt minus3pt%
\belowdisplayskip \abovedisplayskip
\abovedisplayshortskip \z@ plus3pt%
\belowdisplayshortskip 7pt plus3.5pt minus0pt
\def\@listi{\parsep 4.5pt plus 2pt minus 1pt
     \itemsep \parsep
     \topsep 9pt plus 3pt minus 3pt}}

\relax

\catcode`@=12

\catcode`\@=11

\def\section{\@startsection{section}{1}{\z@}{3.5ex plus 1ex minus
   .2ex}{2.3ex plus .2ex}{\large\bf}}

\def\SymBoxes#1#2#3#4{\newdimen\un@t \un@t#3%
\raisebox{#1}{\rule{#2\un@t}{#4}\hskip-#2\un@t
\@tempdimb\un@t \advance\@tempdimb by-#4\@tempcntb#2\relax%
\@whilenum{\@tempcntb>0}\do{
\rule{#4}{\un@t}\hskip\@tempdimb \advance\@tempcntb by\m@ne}%
\hskip-#2\un@t \rule[\un@t]{#2\un@t}{#4}%
\rule[\un@t]{#4}{#4}\hskip-#4
\rule{#4}{\un@t}}\hskip-#4}                

\begin{document}

\newcommand{\beq}{\begin{equation}}
\newcommand{\eeq}{\end{equation}}
\newcommand{\bea}{\begin{eqnarray}}
\newcommand{\eea}{\end{eqnarray}}
\newcommand{\beas}{\begin{eqnarray*}}
\newcommand{\eeas}{\end{eqnarray*}}
\newcommand{\defi}{\stackrel{\rm def}{=}}
\newcommand{\non}{\nonumber}
\newcommand{\bquo}{\begin{quote}}
\newcommand{\enqu}{\end{quote}}
\renewcommand{\(}{\begin{equation}}
\renewcommand{\)}{\end{equation}}
\def \eqn#1#2{\begin{equation}#2\label{#1}\end{equation}}

\def\e{\epsilon}
\def\IZ{{\mathbb Z}}
\def\IR{{\mathbb R}}
\def\IC{{\mathbb C}}
\def\IQ{{\mathbb Q}}
\def\de{\partial}
\def\Tr{ \hbox{\rm Tr}}
\def\H{ \hbox{\rm H}}
\def\HE{ \hbox{$\rm H^{even}$}}
\def\HO{ \hbox{$\rm H^{odd}$}}
\def\K{ \hbox{\rm K}}
\def\Im{ \hbox{\rm Im}}
\def\Ker{ \hbox{\rm Ker}}
\def\const{\hbox {\rm const.}}
\def\o{\over}
\def\im{\hbox{\rm Im}}
\def\re{\hbox{\rm Re}}
\def\bra{\langle}\def\ket{\rangle}
\def\Arg{\hbox {\rm Arg}}
\def\Re{\hbox {\rm Re}}
\def\Im{\hbox {\rm Im}}
\def\exo{\hbox {\rm exp}}
\def\diag{\hbox{\rm diag}}
\def\longvert{{\rule[-2mm]{0.1mm}{7mm}}\,}
\def\a{\alpha}
\def\dag{{}^{\dagger}}
\def\tq{{\widetilde q}}
\def\p{{}^{\prime}}
\def\W{W}
\def\N{{\cal N}}
\def\hsp{,\hspace{.7cm}}

\def\br{\nonumber}
\def\IZ{{\mathbb Z}}
\def\IR{{\mathbb R}}
\def\IC{{\mathbb C}}
\def\IQ{{\mathbb Q}}
\def\IP{{\mathbb P}}
\def \eqn#1#2{\begin{equation}#2\label{#1}\end{equation}}

\newcommand{\C}{\ensuremath{\mathbb C}}
\newcommand{\Z}{\ensuremath{\mathbb Z}}
\newcommand{\R}{\ensuremath{\mathbb R}}
\newcommand{\rp}{\ensuremath{\mathbb {RP}}}
\newcommand{\cp}{\ensuremath{\mathbb {CP}}}
\newcommand{\vac}{\ensuremath{|0\rangle}}
\newcommand{\vact}{\ensuremath{|00\rangle}                    }
\newcommand{\oc}{\ensuremath{\overline{c}}}
\newcommand{\psizero}{\psi_{0}}
\newcommand{\phizero}{\phi_{0}}
\newcommand{\hzero}{h_{0}}
\newcommand{\psiin}{\psi_{\rh}}
\newcommand{\phiin}{\phi_{\rh}}
\newcommand{\hin}{h_{\rh}}
\newcommand{\rh}{r_{h}}
\newcommand{\rb}{r_{b}}
\newcommand{\psibnd}{\psi_{0}^{b}}
\newcommand{\psibndp}{\psi_{1}^{b}}
\newcommand{\phibnd}{\phi_{0}^{b}}
\newcommand{\phibndp}{\phi_{1}^{b}}
\newcommand{\gbnd}{g_{0}^{b}}
\newcommand{\hbnd}{h_{0}^{b}}
\newcommand{\zh}{z_{h}}
\newcommand{\zb}{z_{b}}
\newcommand{\man}{\mathcal{M}}
\newcommand{\hbr}{\bar{h}}
\newcommand{\tbr}{\bar{t}}

\begin{titlepage}
\begin{flushright}
CHEP XXXXX
\end{flushright}
\bigskip
\def\thefootnote{\fnsymbol{footnote}}

\begin{center}
{\Large
{\bf Machine Learning ${\cal N}=8, D=5$ Gauged Supergravity
}
}
\end{center}

\bigskip
\begin{center}
Chethan KRISHNAN$^a$\footnote{\texttt{chethan.krishnan@gmail.com}}, \ Vyshnav MOHAN$^a$\footnote{\texttt{vyshnav.vijay.mohan@gmail.com}}, \ Soham RAY$^a$\footnote{\texttt{raysoham14@gmail.com}}  
\vspace{0.1in}

\end{center}

\renewcommand{\thefootnote}{\arabic{footnote}}

\begin{center}

$^a$ {Center for High Energy Physics,\\
Indian Institute of Science, Bangalore 560012, India}\\

\end{center}

\noindent
\begin{center} {\bf Abstract} \end{center}
Type IIB string theory on a 5-sphere gives rise to ${\cal N}=8, SO(6)$ gauged supergravity in five dimensions.  Motivated by the fact that this is the context of the most widely studied example of the AdS/CFT correspondence, we undertake an investigation of its critical points. The scalar manifold is an $E_{6(6)}/USp(8)$ coset, and the challenge is that it is 42-dimensional. We take a Machine Learning approach to the problem using TensorFlow, and this results in a substantial increase in the number of known critical points. Our list of 32 critical points contains all five of the previously known ones, including an ${\cal N}=2$ supersymmetric point identified by Khavaev, Pilch and Warner.



\vspace{1.6 cm}
\vfill

\end{titlepage}

\setcounter{footnote}{0}

\section{Introduction}

The most intensely studied example of the AdS/CFT correspondence is that between type IIB string theory on AdS$_5 \times S^5$ supported by $N$ units of 5-form flux, and ${\cal N}=4$ superconformal $SU(N)$ Yang-Mills theory in four dimensions \cite{Malda}. It is generally believed that the 10-dimensional theory allows a consistent truncation to five dimensions\footnote{See \cite{consistent} for discussions on consistent truncation to gauged maximal supergravity in various dimensions. More recently, exceptional field theory has been used to relate the 10 and 5D languages in  \cite{Hohm} and argue conclusively that the IIB truncation to 5D maximal gauged SUGRA is consistent.}. This means that one can restrict one's attention to a finite number of five dimensional fields including the metric, and they do not couple to the rest of the (otherwise infinite) number of fields that arise in five dimensions (including the higher Kaluza-Klein harmonics). The resulting theory is the gauged ${\cal N}=8$ supergravity in five dimensions \cite{Gunaydin0, Pernici, Gunaydin, Samt}. From the dual gauge theory perspective, these finite number of consistently truncated fields are dual to the short ${\cal N}=4$ multiplet containing the energy momentum tensor, and the statement of consistent truncation translates to the statement that at least at large-$N$, this particular class of chiral primary operators close under the Operator Product Expansion (OPE). Since the supergravity scalars capture the relevant and marginal couplings, the fact that there is a consistent truncation suggests that the renormalization group (RG) flow triggered by them in the gauge theory can be fully captured in the supergravity. In particular, the vacua of the gauged supergravity should capture the IR fixed points of such large-$N$ flows. Therefore, understanding the vacuum structure of these supergravities is of interest from multiple perspectives.

Finding the vacua of gauged supergravities is a conceptually trivial problem: one just has to find the extremal points of the scalar potential in this theory. The vacua of ${\cal N}=8, D=5$ gauged supergravity have negative vacuum energy and correspond to AdS vacua, and because of the Breitenlohner-Freedman criterion \cite{BF}, they may be stable even when they are extrema and not necessarily minima. Despite the conceptual simplicity of the problem however, only a handful of vacua have been identified in the nearly  35 years since the discovery of the theory in \cite{Gunaydin0, Pernici, Gunaydin}. The trouble here is two-fold. Firstly, the number of scalars in these theories is large. ${\cal N}=8, D=5$ gauged supergravity has 42 scalars, and even if we were to somehow take advantage of the fact that the potential has an $SU(1,1) \times SO(6)$ symmetry, the number of scalars would still be 24. Secondly, the potential is complicated and has a baroque structure arising from the underlying gauging of the theory. 

Due to these facts, a systematic effort at finding the critical points of ${\cal N}=8$, $D=5$ gauged supergravity has not been undertaken to the best of our knowledge. In the literature, we are aware of five distinct critical points. The first three (including the maximally supersymmetric one) were noted already at the time when the theory was constructed \cite{Gunaydin}. Later, after the advent of the AdS/CFT correspondence, a further two were identified by Khavaev, Pilch and Warner \cite{Khavaev} in a certain $SU(2)$ invariant subsector of the theory. One of these two new critical points has ${\cal N}=2$ supersymmetry. The situation has been better for the maximal gauged supergravity in four dimensions, where steady effort has been put forth by many authors, in particular by Fischbacher and collaborators in the last decade or so (see eg. \cite{Fisch}), to identify quite a few vacua. 

In this paper, we will take advantage of Machine Learning (ML) as a recently emerging tool for finding vacua of string theory (see references to \cite{Carifio:2017bov, Bull:2018uow} for related examples).  An ML approach using TensorFlow was adopted in \cite{Fischbacher} (see also \cite{Bobev1}) to find critical points of ${\cal N}=8$, $D=4$ supergravity \cite{dWN}, and our approach will be closely parallel. We will find that this approach is surprisingly powerful, and will be able to go (much) beyond the five previously known critical points. In particular, we will find 32 critical points including all the five previously known ones. We suspect that this list is exhaustive, but we should warn the reader that some of the critical points are found exceedingly more rarely ($\sim$ 500:1) than some others during the ML search, so it is difficult to rule out the possibility that we have missed some ``rare" critical point(s). 

Most of our discussion in what follows has to do with technicalities in the implementation of the gauge/global symmetries  and the coset structure in a form suitable for direct calculations with TensorFlow. So most of this paper is to be found in the Appendices. In the main text, we will simply quote the action of the theory, discuss the previously known critical points to give some analytic context for the problem, and then summarize some of the salient features of our approach and the new critical points we find. 

{\bf Note Added:} The paper \cite{Bobev2} that appeared after our work, also addresses the question of critical points of ${\cal N}=8, D=5$ gauged supergravity using TensorFlow. It has become clear from  their correspondence with us in the last week, that except for the fact that they do not find our critical point \#26, our results agree in great detail. This is a very strong Bayesian check that these results are indeed correct. 

Let us make a few comments about the validity of \#26. In our numerics, it has the same robustness as the other critical points: we have found this critical point using both loss functions and when scanning for both 42 or 24 scalars (see discussions later). The tolerance of the loss functions is about $\sim 10^{-23}$. Since the scalar values and potential values are ${\cal O}(1)$ numbers, this hierarchy is best explained by a zero. Note also that finding the critical point is the hard part. Checking its validity once its location is known is trivial, and we find results consistent with  the loss function tolerance. Because of these reasons, we believe \#26 is a legitimate critical point.


\section{The Action}



The ungauged $\mathcal{N}=8$ supergravity in five dimensions can be obtained via the dimensional reduction of eleven dimensional supergravity. The theory contains one graviton, 8 gravitini $\psi$, 27 vector fields $A_\mu$, $48$ spin-$\frac{1}{2}$ fermions $\chi$ and $42$ scalars $\phi$.  We will be interested in the gauged theory, and its field content is closely related \cite{Gunaydin}. 

To make some of our comments, it will be useful to have the form of the Lagrangian and therefore we will present it below. Defining all of the notation at this stage will be too much of a distraction, we will discuss what we need in later sections and appendices. The reader should also consult \cite{Gunaydin} whose notations we follow. We will write our expressions in a form where the gauging is $SO(p,6 - p)$ with $p = 0,1,2$ or $3$. The choice of $p$ is reflected in the signature of $\eta^{IJ}$. We will mostly be concerned with the  $p=0$ case, which is the case that has immediate connections with string theory. Excluding four fermion terms, the Lagrangian has the form
\begin{equation}
\begin{aligned}
e^{-1} \mathcal{L}=&-\frac{1}{4} R-\frac{1}{2} i \bar{\psi}_{\mu}^{a} \gamma^{\mu \nu \rho} D_{\nu} \psi_{\rho a}+\frac{1}{12} i \bar{\chi}^{a b c} \gamma^{\mu} D_{\mu} \chi_{a b c} \\
&+\frac{1}{24} P_{\mu a b c d} P^{\mu a b c d}-\frac{1}{8} H_{\mu \nu a b} H^{\mu \nu a b}+\frac{1}{3} \sqrt{\frac{1}{2}} i P_{\nu a b c d} \bar{\psi}_{\mu}^{a} \gamma^{\nu} \gamma^{\mu} \chi^{b c d} \\
&+\frac{1}{4} i H_{\mu \nu}^{a b}\left[\bar{\psi}_{a}^{\rho} \gamma_{[ \rho} \gamma^{\mu \nu} \gamma_{\sigma ]} \psi_{b}^{\sigma}+\sqrt{\frac{1}{2}} \bar{\psi}_{\rho}^{c} \gamma^{\mu \nu} \gamma^{\rho} \chi_{a b c}+\frac{1}{2} \bar{\chi}_{a c d} \gamma^{\mu \nu} \chi_{b}^{c d}\right] \\
&-\frac{1}{15} i g T_{a b} \bar{\psi}_{\mu}^{a} \gamma^{\mu \nu} \psi_{\nu}^{b}+\frac{1}{6} \sqrt{\frac{1}{2}} i g A_{d a b c} \bar{\chi}^{a b c} \gamma^{\mu} \psi_{\mu}^{d} +\frac{1}{2} ig \bar{\chi}^{a b c}\left(\frac{1}{2} A_{b c d e}-\frac{1}{45} \Omega_{b d} T_{c e}\right) \chi_{a}^{d e} \\
&+\frac{1}{96} g^{2}\left[\frac{64}{225}\left(T_{a b}\right)^{2}-\left(A_{a b c d}\right)^{2}\right] \\
&-\frac{1}{96 e} \varepsilon^{\mu \nu \rho \sigma \tau} \varepsilon^{I J K L M N}\Big[ F_{I J \mu \nu} F_{K L \rho \sigma} A_{M N \tau}+g \eta^{P Q} F_{I J \mu \nu} A_{K L \rho} A_{M P\sigma} A_{Q N \tau}\\
&+\frac{2}{5} g^{2} \eta^{P Q} \eta^{R S} A_{I J \mu} A_{K P r} A_{Q L \rho} A_{M R \sigma} A_{S N \tau}\Big] \\
&+\frac{1}{8 g e} \varepsilon^{\mu \nu \rho \sigma \tau} \eta_{I J} \varepsilon_{\alpha \beta} B_{\mu \nu}^{I \alpha} D_{\rho} B_{\sigma \tau}^{J \beta}.
\end{aligned} \label{lagrangian}
\end{equation}

\noindent
Our primary focus will be on the scalar term quadratic in the coupling constant $g$. This corresponds to the potential of the theory.

From the Lagrangian, we can see that the masses of the gravitini depend on the term $\sim g T\psi  \bar{\psi}$. These masses will be helpful in determining the calculating the residual supersymmetry at a critical point. We define the ``naive'' gravitino masses $m^{2}/m_{0}^{2}(\psi)$ to be the eigenvalues of the matrix $M_\psi$, given by
\bea
M_{\psi} =\frac{L^{2}}{225} T_{ab}T^{ac} = -\frac{10}{P_0} \frac{1}{225} T_{ab}T^{ac} \label{gravitino}
\eea
with $m_0^2(\psi)\equiv L^2/225$. We have normalized masses using the AdS radius $L^2 = -D(D-1)/2P_0$, where $D=5$ is the dimensionality of spacetime and $P_0$ is the potential at the critical point. This normalization ensures that the unbroken supersymmetric critical points have a naive gravitino mass of $+1$. 

\section{The Scalar Potential}
\label{potentialsection}

The basic ingredient which goes into the construction of the potential is the vielbein. They characterize the scalar manifold, which is a coset $E_{6(6)}/USp(8)$ whose details are presented in appendices. The vielbein is written as $V_{AB}{}^{ab}$ and its definition and properties are given in Appendix \ref{uspappendix}. Using the vielbein, we can construct certain objects called $T$-tensors, which play a crucial role in constructing the potential. 

In a certain $SL(6, \mathbb{R}) \times SL(2, \mathbb{R})$ basis, the vielbein can be broken down into (see Appendix \ref{sl6appendix})
\bea
V_{AB}{}^{ab} = (V^{IJab},V_{I\alpha}{}^{ab})
\eea
The Roman indices $I,J,K$ run from 1 to 6 while the Greek indices $\alpha,\beta$ take values 1 and 2. Using this splitting of the vielbein, and defining a tensor $W_{abcd}$:
\bea
W_{abcd} = \varepsilon^{\alpha\beta}\eta^{IJ}V_{I\alpha ab}V_{J\beta cd}
\eea
where $\varepsilon^{\alpha\beta}$ is the 2-dimensional Levi-Civita tensor. $\eta^{IJ}$ is the unit diagonal matrix with signature $(p,6-p)$ and encodes the details of the $SO(p,6-p)$ gauging. The $W_{abcd}$ tensor has the property that
\bea
W_{abcd} = -W_{cdab}
\eea
This lets us define a symmetric tensor $T_{ab}$ as (the symplectic form $\Omega$ is defined in the appendices)
\bea
T_{ab} = -\frac{15}{4}\Omega^{cd}W_{acbd}
\eea
We also define the tensor $A_{abcd}$ as 
\bea
A_{abcd} = -3 W_{a[bcd]|}
\eea
The vertical bar subscript indicates that we are subtracting out the symplectic trace. We antisymmetrize the last 3 indices and remove the symplectic traces corresponding to those in order to define $A_{abcd}$. This means that:
\bea
\Omega^{bc}A_{abcd} = \Omega^{cd}A_{abcd} = \Omega^{bd}A_{abcd} = 0
\eea

\noindent
Now, having defined these objects (which can be called $T$-tensors) we can go on to the expression for the potential. The potential, in terms of $T_{ab}$, $A_{abcd}$ and the coupling constant $g$ in the Lagrangian is:
\bea
P = -g^{2} \left[\frac{6}{45^2} (T_{ab})^{2} - \frac{1}{96} (A_{abcd})^{2}\right]
\eea

This is the order $g^2$ scalar term in the Lagrangian \eqref{lagrangian}. An equivalent expression for the potential is given by:
\bea
P = -\frac{1}{32}g^{2}[2(W_{ab})^{2} - (W_{abcd})^{2}]
\eea
\noindent
where $W_{ab}$ is defined as:
\bea
W_{ab} = \frac{4}{15} T_{ab}
\eea

\section{Old Critical Points}

As mentioned in the introduction, 5 of the 32 critical points we find were known before: we will call them {\em pre-historic} \cite{Gunaydin} and {\em ancient} \cite{Khavaev} critical points. We will describe them here, both as a way to give some context and also as a way to gain some intuition on the analytic aspects. The $SL(6, \mathbb{R}) \times SL(2, \mathbb{R})$ subgroup structure of these critical points is instructive more broadly. 

When all the scalars parametrizing the potential vanish, we have the trivial critical point, with potential value $-\frac{3}{4}g^2$. This is the maximally supersymmetric point, with $\mathcal{N} =8$. For a critical point with potential $P_0$ its supersymmetry is given by the number of eigenvalues $\mu$ of $W_{ab}$ evaluated at that point satisfying \cite{Gunaydin}:
\bea
|\mu| = \sqrt{-3 P_0/g^{2} } \label{susy}
\eea
\noindent
All the 8 eigenvalues of $W_{ab}$ at the trivial critical point have value $-3/2$, and thus this point is maximally supersymmetric.

To discuss the other critical points, we need a bit more technology. From Appendix \ref{exponentiatingappendix}, we can see that in the $SL(6, \mathbb{R}) \times SL(2, \mathbb{R})$ basis, the vielbein is defined in terms of the four blocks of a $27 \times 27$ matrix $U$, which we denote by $U^{MN}{}_{IJ}$, $U^{IJK\alpha}$, $U_{P\beta IJ}$ and $U_{P\beta}{}^{K\alpha}$. We consider the $SL(6, \mathbb{R}) \times SL(2, \mathbb{R})$ sector of the scalar manifold, in which we have:
\bea
\begin{aligned}
U^{MN}{}_{IJ} &= 2 S_{[I}{}^{[M}S_{J]}{}^{N]} \\
U^{IJK\alpha} &= U_{K\alpha IJ} = 0 \\
U_{I\beta}{}^{J\alpha} &= S_{I}{}^{J} S'_{\beta}{}^{\alpha}.
\end{aligned}
\eea
Here $S$ is an $SL(6, \mathbb{R})$ matrix while $S'$ is an $SL(2, \mathbb{R})$ matrix. In terms of the symmetric matrix $M_{IJ} = S_{I}{}^{K}S_{J}{}^{K}$, we have:
\bea
W_{ab} = -\frac{1}{4} \text{Tr}(M) \delta_{ab}
\eea
The potential in terms of $M$ takes the form (we are restricting ourselves to the $SO(6)$ gauging):
\bea
P = -\frac{1}{32} g^{2} \{ [\text{Tr}(M)]^{2} - 2 \text{Tr}(M^{2}) \}  
\eea

Now we are ready to discuss another pre-historic critical point. We pick an $SO(5)$ invariant direction by picking:
\bea
M = \diag({e^{\lambda},e^{\lambda},e^{\lambda},e^{\lambda},e^{\lambda},e^{-5\lambda}})
\eea
The potential in this sector now becomes:
\bea
P = -\frac{1}{32} g^{2} \{15e^{2\lambda} + 10e^{-4\lambda} - e^{-10\lambda} \}
\eea
This potential has 2 critical points. The first one, at 
$\lambda = 0$ is just the ${\cal N}=8$ critical point again. The second critical point occurs at $\lambda = -\frac{1}{6}$ log(3). This is an $SO(5)$ invariant critical point with potential:
\bea
P_0 = -\frac{3^{5/3}}{8} g^2
\eea
All the eigenvalues of $W_{ab}$ at this point are equal to $-\frac{2}{3^{1/6}}$, hence it has no supersymmetry. This point was shown to be perturbatively unstable in \cite{Distler}.

We now search for the $SU(3)$ invariant critical points in the theory. We start by defining an anti-self-dual $SO(8)$ tensor $X_{ijkl}$ as:
\bea
X_{ijkl} = -(\delta_{ijkl}^{1357}-\delta_{ijkl}^{2468})+(\delta_{ijkl}^{1368}-\delta_{ijkl}^{2457})+(\delta_{ijkl}^{1458}-\delta_{ijkl}^{2367})+(\delta_{ijkl}^{1467}-\delta_{ijkl}^{2358})
\eea
The indices $i,j,k$ and $l$ run from 1 to 8. Identifying $(\alpha,\beta) = (7,8)$ we define the $SU(3)$ invariant scalar field to be:
\bea
\varphi_{IJK\alpha} = \lambda X_{IJK\alpha}
\eea
Defining $p =$ cosh$(4\lambda)$ we find:
\bea
W_{ab} = -\frac{3}{8}(p+3)\delta_{ab} + \frac{1}{16}(p-1) J_{ab}
\eea
\noindent
where $J_{ab} = \diag (2,2,2,2,2,2,-6,-6)$. The potential is:
\bea
P = \frac{3}{32}g^{2} (p^2 - 4p -5)
\eea
The critical points of the potential are at $p=1$ and $p=2$. The $p=1$ point corresponding to $\lambda=0$ is the maximally supersymmetric point that we have obtained earlier. For $p=2$ we have an $SU(3)$ invariant critical point with potential:
\bea
P_0 = -\frac{27}{32}g^2
\eea
The eigenvalues of $W_{ab}$ here are $-\frac{7}{4}$ and $-\frac{9}{4}$ with multiplicities of 6 and 2 respectively. There is no supersymmetry at this point.

Now we turn to the {\em ancient} critical points discovered in \cite{Khavaev}. There is an $SU(2) \times U(1) \times U(1)$ critical point with potential:
\bea
P_0 = -\frac{3}{8} \left(\frac{25}{2}\right)^{1/3} g^2
\eea
The eigenvalues of $W_{ab}$ at this critical point are $-\frac{3}{10^{1/6}}$ and $-\frac{9}{10^{2/3}}$, both of which have a multiplicity of 4. Thus there is no supersymmetry.

The fifth and final known critical point is an interesting one. This is an $SU(2) \times U(1)$ critical point with potential:
\bea
P_0 = -\frac{2^{4/3}}{3}g^2
\eea
This critical point is generated by 
\bea
\varphi_{IJK\alpha} &=& \frac{1}{4} \log{(3)} \ X_{IJK\alpha} \\
S^{I}{}_{J} &=& \diag (e^{\alpha},e^{\alpha},e^{\alpha},e^{\alpha},e^{-2\alpha},e^{-2\alpha})
\eea
\noindent
where $\alpha = \frac{1}{12}$ log$(2)$. The eigenvalues of $W_{ab}$ are $-\frac{7}{3} 2^{-1/3}$, $-\frac{4}{3} 2^{2/3}$ and $-2^{2/3}$ with multiplicities 4, 2 and 2 respectively. Note that the third eigenvalue satisfies $\mu = -(-\frac{3 P_0}{g^2})^{1/2}$. Thus this point has a supersymmetry of ${\cal N}=2$.

\section{Leveling the Field with TensorFlow}

Historically \cite{Warner}, the most effective way to search for critical points was to restrict the problem to a space of singlets of some invariance group. Schur's lemma guarantees that a critical point on this subspace would be the critical point  of the full scalar manifold. This strategy has been employed in \cite{Khavaev} to find the $5$ critical points which had at least a residual symmetry of $SU(2)$. However this method is not useful when a critical point breaks all the symmetries, for example. 

This is where ML techniques come to our rescue. Google's TensorFlow libraries have been used in \cite{Fischbacher} to find out critical points of the $\mathcal{N} = 8$ $D = 4$ $SO(8)$ gauged supergravity scalar potential. We have followed  the same philosophy in this paper and found that it helps us make substantial progress. The key idea is to reinterpret the $42$ scalars as a set of learnable parameters and then to tune them to minimize an appropriately defined ``loss function''. Appendix \ref{lossfunctionapp} contains the discussion on various loss functions that we have used to find the critical points.  

\begin{table*}\centering

\begin{tabular}{@{} l *4c @{}}

\toprule

 &{\#}   \phantom{abcdefghijkl} &  {Potential} \phantom{abcdefghijkl}  & {$\mathcal{N}$} \phantom{abcdefghijkl}  & {History}\\

\midrule

 &1  \phantom{abcdefghijkl} & $ -0.75 \ g^{2}$ \phantom{abcdefghijkl} & 8  \phantom{abcdefghijkl} &GRW1\\

 &2  \phantom{abcdefghijkl} & $  -0.78003 \ g^{2}$ \phantom{abcdefghijkl} & 0 \phantom{abcdefghijkl} &GRW2\\

 &3  \phantom{abcdefghijkl} & $   -0.83995 \ g^{2}$ \phantom{abcdefghijkl} & 2  \phantom{abcdefghijkl} &KPW2\\

&4  \phantom{abcdefghijkl} & $  -0.84375 \ g^{2}$ \phantom{abcdefghijkl} & 0 \phantom{abcdefghijkl} &GRW3\\

&5  \phantom{abcdefghijkl} & $   -0.870298\ g^{2}$ \phantom{abcdefghijkl} & 0 \phantom{abcdefghijkl}&KPW1\\

&6  \phantom{abcdefghijkl} & $   -0.87894 \ g^{2}$ \phantom{abcdefghijkl} & 0 \phantom{abcdefghijkl} & -\\

&7  \phantom{abcdefghijkl} & $   -0.88764 \ g^{2}$ \phantom{abcdefghijkl} & 0 \phantom{abcdefghijkl} &-\\

&8  \phantom{abcdefghijkl} & $   -0.89291\ g^{2}$ \phantom{abcdefghijkl} & 0 \phantom{abcdefghijkl} &-\\

&9  \phantom{abcdefghijkl} & $  -0.96395 \ g^{2}$ \phantom{abcdefghijkl} & 0 \phantom{abcdefghijkl} &-\\

&10  \phantom{abcdefghijkl} & $  -0.96410 \ g^{2}$ \phantom{abcdefghijkl} & 0 \phantom{abcdefghijkl} &-\\

&11  \phantom{abcdefghijkl} & $   -0.96453 \ g^{2}$ \phantom{abcdefghijkl} & 0 \phantom{abcdefghijkl} &- \\

&12  \phantom{abcdefghijkl} & $   -0.98278 \ g^{2}$ \phantom{abcdefghijkl} & 0 \phantom{abcdefghijkl} &-\\

&13  \phantom{abcdefghijkl} & $   -1.00148 \ g^{2}$ \phantom{abcdefghijkl} & 0 \phantom{abcdefghijkl} &-\\

&14  \phantom{abcdefghijkl} & $  -1.05469 \ g^{2}$ \phantom{abcdefghijkl} & 0 \phantom{abcdefghijkl} &-\\

&15  \phantom{abcdefghijkl} & $   -1.07353\ g^{2}$ \phantom{abcdefghijkl} & 0 \phantom{abcdefghijkl} &-\\

&16  \phantom{abcdefghijkl} & $   -1.125 \ g^{2}$ \phantom{abcdefghijkl} & 0 \phantom{abcdefghijkl} &-\\

&17  \phantom{abcdefghijkl} & $   -1.29725 \ g^{2}$ \phantom{abcdefghijkl} & 0 \phantom{abcdefghijkl} &-\\

&18  \phantom{abcdefghijkl} & $  -1.30291 \ g^{2}$ \phantom{abcdefghijkl} & 0 \phantom{abcdefghijkl} &-\\

&19  \phantom{abcdefghijkl} & $  -1.30461 \ g^{2}$ \phantom{abcdefghijkl} & 0 \phantom{abcdefghijkl} &-\\

&20  \phantom{abcdefghijkl} & $  -1.31918 \ g^{2}$ \phantom{abcdefghijkl} & 0 \phantom{abcdefghijkl} &-\\

&21  \phantom{abcdefghijkl} & $   -1.38225 \ g^{2}$ \phantom{abcdefghijkl} & 0 \phantom{abcdefghijkl} &-\\

&22  \phantom{abcdefghijkl} & $  -1.39104 \ g^{2}$ \phantom{abcdefghijkl} & 0 \phantom{abcdefghijkl} &-\\

&23  \phantom{abcdefghijkl} & $  -1.416746 \ g^{2}$ \phantom{abcdefghijkl} & 0 \phantom{abcdefghijkl} &-\\

&24  \phantom{abcdefghijkl} & $  -1.417411 \ g^{2}$ \phantom{abcdefghijkl} & 0 \phantom{abcdefghijkl} &-\\

&25  \phantom{abcdefghijkl} & $  -1.460654 \ g^{2}$ \phantom{abcdefghijkl} & 0 \phantom{abcdefghijkl} &-\\

&26  \phantom{abcdefghijkl} & $  -1.460730 \ g^{2}$ \phantom{abcdefghijkl} & 0 \phantom{abcdefghijkl} &-\\

&27  \phantom{abcdefghijkl} & $  -1.497042 \ g^{2}$ \phantom{abcdefghijkl} & 0 \phantom{abcdefghijkl} &-\\

&28  \phantom{abcdefghijkl} & $  -1.499667 \ g^{2}$ \phantom{abcdefghijkl} & 0 \phantom{abcdefghijkl}&-\\

&29  \phantom{abcdefghijkl} & $  -1.501862 \ g^{2}$ \phantom{abcdefghijkl} & 0 \phantom{abcdefghijkl} &-\\

&30  \phantom{abcdefghijkl} & $  -1.510901  \ g^{2}$ \phantom{abcdefghijkl} & 0 \phantom{abcdefghijkl}&-\\

&31  \phantom{abcdefghijkl} & $  -1.547778  \ g^{2}$ \phantom{abcdefghijkl} & 0 \phantom{abcdefghijkl}&- \\

&32  \phantom{abcdefghijkl} & $  -1.73841 \ g^{2}$ \phantom{abcdefghijkl} & 0 \phantom{abcdefghijkl}&-\\ \bottomrule
 
\end{tabular}

\caption{The list of all critical points. The second column contains the value of the potential at the critical point and the third column counts the unbroken SUSY. The final column refers to the original paper where the critical point first appeared (GRW refers to \cite{Gunaydin}, KPW to \cite{Khavaev}), with the numbering indicating the sequence in which it appears in our discussion in section 4. }

\label{table1}

\end{table*}

\subsection{The Choice of Scalars in the Vielbein}

The most essential ingredient in constructing the potential is the vielbein (See Appendix \ref{uspappendix} for a discussion on the properties of the vielbein). The vielbein captures an element of the $E_{6(6)}/USp(8)$ coset, and to construct it we need to use the Lie algebra of $E_{6(6)}$ in a suitable basis and identify the non-compact part of the algebra that generates the coset.

As we are looking at gaugings of the type $SO(p,6-p)$, it is convenient to write the Lie algebra in the $SL(6, \mathbb{R}) \times SL(2, \mathbb{R})$ basis, as done in \cite{Gunaydin}. From Appendix \ref{sl6appendix}, we can see that the Lie algebra has three sets of generators in this basis. Using the basis elements defined in Appendix \ref{explicitbasisapp}, we can construct the $E_{6(6)}/USp(8)$ generators as 
\bea
{S^{K}}_{L} = \phi^{IJ}{\left(S_{IJ}\right)^{K}}_{L} 
\eea
\bea
{S^{\gamma}}_{\sigma} = \phi^{\alpha \beta}{\left(S_{\alpha \beta}\right)^{\gamma}}_{\sigma} 
\eea
\bea
\Sigma^{+}_{I J K \alpha} =\phi^{PQR \beta}(\Sigma^{+}_{PQR \beta})_{I J K \alpha}  
\eea
where we have introduced a set of $42$ scalars $(\phi^{IJ}, \phi^{\alpha \beta},\phi^{PQR \beta})$ to contract the generator indices of the basis elements. The vielbein can be constructed in terms of these $42$ scalars using the set of formulae given in Appendix \ref{exponentiatingappendix}. The values of these scalars at the critical points are what we report in our Appendix  \ref{criticalpointappendix}.

\subsection{Loss Function Minimization}

Once we have the vielbein, the construction of the potential and the loss function is quite straightforward. The next step is to minimize the loss function. Here is where we will rely on Machine Learning by using Google's TensorFlow library on Google colab \cite{colab}. Strictly speaking one can also install and run TensorFlow locally on a Python environment (and we have), but Google colab enables us to bypass local system limitations by relying on cloud computing.

To minimize the loss function, we start off on a random location on the scalar manifold by drawing random samples from a Gaussian distribution. The mean of the distribution is obtained from a pseudo-random number generator. Therefore, we have two adjustable parameters that determine the value of the scalars - the key of the pseudorandom number generator, called \textit{seed} and the standard deviation of the distribution, called \textit{scale}. 

For some specific value of the seed and the scale, the loss function can be minimized by using the helper function \code{ tf.contrib.opt.ScipyOptimizerInterface()}. This was done in \cite{Fischbacher} and we will adapt their code. By changing the value of the seed and scale systematically, we can scan various regions of the scalar manifold to find critical points of the potential. The precise way in which this is most optimally done requires a bit of trial and error. Changing seeds for a fixed ``large" value of the scale (``large" here can be thought of as an ${\cal O}(1)$ number) has worked well for us in retrieving all the critical points. 

We have done the ML search with multiple loss functions (see Appendix \ref{lossfunctionapp}), and with and without explicitly fixing the global symmetry $SU(1,1) \times SO(6)$ of the potential. The results we find for the critical points, are stable. Our numerical results exactly match the values of the old critical points up to the precision we have looked at\footnote{After v1 of this paper appeared, T. Fischbacher has contacted us and informed us that he and collaborators have also made progress on this problem. He has sent us some of the critical points. They agree with what we had reported, to the number of decimals he has shown us. This may be viewed as another argument for believing that our results are correct.}. 

There are two sets  of two critical points (see Table) that differ in their value only at the forth decimal. When we are running the code at high precision, all the loss function values are between  $\sim 10^{-20}$ and $\sim 10^{-30}$, so we are confident that they are distinct and that this is not a numerical artifact. Another check of this is that they have distinct gravitino masses. Since some of the critical points that we obtained are closely spaced, it is necessary to tighten various tolerance values in the code. This has to be done simultaneously with a change in the internal parameters of the L-BFGS-B algorithm that we are using in our helper function. This is because otherwise, the algorithm might exit the minimization procedure even before the required level of tolerance is reached.  

\subsection{Discussion}

After solving about $100$ $000$ numerical minimization problems, we have obtained $32$ distinct critical points. This includes $27$ new critical points that were not known before (Refer Table \ref{table1} for the complete list.). The gravitino masses $m^{2}/m_{0}^{2}(\psi)$ and the location of these critical points on the scalar manifold have been given in the Appendix \ref{criticalpointappendix}. 

We plan to give a detailed analysis of the properties of these critical points in an upcoming paper.  Unlike the ${\cal N}=8, D=4$ case where much work has been done on various aspects, investigations on the critical points of  ${\cal N}=8$, $D=5$ theory seem sparse. It will be most useful to express the scalar and fermion mass matrices while paying heed to the residual symmetries of the critical point. This will also be of interest in studying the BF stability of these critical points. 

We have also done a preliminary scan of critical points in some of the non-compact gaugings - this reproduces the rudimentary results mentioned in \cite{Gunaydin}. It is perhaps worth undertaking a more intense effort in this direction, but since the question of unitary completions of the non-compact gaugings is less clear, we will not do so here.

\section{Acknowledgments}

We thank Thomas Deppisch for a correspondence on $E_6$ and Prasad Hegde and Venkata Lokesh Kumar for discussions on Machine Learning.   

\appendix

\section{$E_6$, $USp(8)$ and $E_{6(6)}/USp(8)$}
\label{uspappendix}
\noindent
To study the properties of the $E_{6(6)}/USp(8)$ coset of the gauged theory, it is most convenient to write the Lie algebra in a particular basis, called the $USp(8)$ basis.  The group $E_{6}$ has dimension 78. The parenthetical $(6)$ in the notation $E_{6(6)}$ is supposed to indicate that the difference between the number of non-compact and compact generators is 6. In other words, we have 42 non-compact and 36 compact generators. The maximal compact subgroup of $E_{6(6)}$ is $USp(8)$. In a basis where the $USp(8)$ structure is manifest (which we will call the $USp(8)$ basis), the 42 non compact directions are generated by $\Sigma_{abcd}$ while the compact directions are generated by $\Lambda^{a}{}_{b}$. The $\Sigma_{abcd}$ generators are fully antisymmetric, symplectic traceless and pseudoreal. The generators $\Lambda^{a}{}_{b}$ are anti-Hermitian and symmetric and straightforwardly constructed. We will not need them here. Together, this completes the definition of the $E_{6(6)}$ algebra.

To define things a bit more concretely, it is useful to define a symplectic antisymmetric matrix $\Omega$ with the properties:
\bea
\Omega^{ab} = (\Omega_{ba})^{*}, \ \ 
\Omega_{ab}\Omega^{bc} = \delta_{a}^{c}
\eea
$\Omega$ is used for raising and lowering of indices:
\bea
X^{a} = \Omega^{ab}X_{b}, \ \
X_{a} = \Omega_{ab}X^{b}
\eea

In terms of the $\Omega$ matrices, we will define a representation for the $\Sigma$ generators as \cite{Sezgin:1981ac}:
\bea
(\Sigma_{abcd})_{ef}{}^{gh} = \Omega_{ei}\Omega_{fj}(\delta_{[a}^{[i}\delta_{b}^{j}\delta_{c}^{g}\delta_{d]}^{h]} - \frac{3}{2}\Omega_{[ab}\delta_{c}^{[i}\delta_{d]}^{j}\Omega^{gh]} + \frac{1}{8}\Omega_{[ab}\Omega_{cd]}\Omega^{[ij}\Omega^{gh]})
\eea
In the next paragraphs we discuss the vector space on which this representation can act and some related properties. 

Note here that $\Sigma$ carries 8 indices. The 4 indices inside the parentheses are the generator indices, and the indices outside the parentheses are the matrix indices. In the rest of this text, whenever $\Sigma$ is written with 4 indices, those are to be understood as matrix indices, unless explicitly stated otherwise. By a generator, we will often mean a linear combination of all the generators. All indices here run from 1 to 8. 

We have already stated that there are 42 generators corresponding to the $E_{6(6)}/USp(8)$ coset, but from the antisymmetry of the generator indices there appear to be 70 ($=  \ ^{8}C_{4}$) independent generators. However the property of symplectic tracing imposes 28 constraints on these generators, leaving us with 42 independent $\Sigma$. The symplectic trace of a 4-index object is defined as $\Omega^{AB}X_{ABCD}$. Thus, the symplectic tracelessness of $\Sigma$ is written as:
\bea
\Omega^{AB}\Sigma_{ABCD} = 0
\eea
\noindent
Note that here $A,B,C,D$ are generator indices. $\Sigma$ also has the property of pseudoreality:
\bea
\Sigma^{ABCD} = (\Sigma_{ABCD})^{*}
\eea
\noindent
$\Sigma^{ABCD}$ and $\Sigma_{ABCD}$ can be obtained from each other by raising and lowering indices with $\Omega$ (while adjusting the matrix indices suitably).

Now let us consider the infinitesimal $E_{6(6)}$ transformations to understand the action of $\Sigma$ on the underlying vector space. The explicit representation we have defined above is in fact the fundamental representation of $E_{6(6)}$. It is real and 27-dimensional. This 27-dimensional vector space can be given a basis $z^{AB}$ with the following properties (with all indices from 1 to 8):
\bea
z^{AB} = -z^{BA}, \ \ 
\Omega_{AB}z^{AB} = 0, \ \ z_{AB} = (z^{AB})^{*}
\eea
The infinitesimal $E_{6(6)}$ transformations are written as:
\bea
\delta z^{AB} = \Lambda^{A}{}_{C}z^{CB} + \Lambda^{B}{}_{C}z^{AC} + \Sigma^{ABCD}z_{CD}
\eea
Let us emphasize that we are working with matrix indices now. 
The $\Lambda$ are the anti-Hermitian tensors which generate $USp(8)$. The above expression can be exponentiated to yield the finite transformations of $E_{6(6)}$:
\bea
(z')^{ab} = V_{AB}^{ab}z^{AB} \label{vielbeineq}
\eea
\noindent
Similar infinitesimal and finite transformations can be written for the conjugate representation of $E_{6(6)}$ which is also 27-dimensional and has basis element $\tilde{z}^{AB}$ satisfying the same properties as $z^{AB}$ but having the transformation:
\bea
\delta \tilde{z}^{AB} = \Lambda^{A}{}_{C}\tilde{z}^{CB} + \Lambda^{B}{}_{C}\tilde{z}^{AC} - \Sigma^{ABCD}\tilde{z}_{CD}
\eea
This in turn leads to 
\bea
\tilde{(z')}_{ab} = \tilde{V}^{AB}_{ab}\tilde{z}_{AB}
\eea
\noindent
The matrices $V$ and $\tilde{V}$ are the vielbeins. As they characterize the action of $E_{6(6)}$ on a 27-dimensional vector space they are called 27-bein. Since we are only interested in the 42-dimensional coset manifold we can gauge away the $USp(8)$ and set its generators $\Lambda$ to zero. Thus, in the $USp(8)$ gauge, the vielbein can be written as the exponential of the $\Sigma$ generators. 

The matrices $V$ and $\tilde{V}$ satisfy property:
\bea
\tilde{V}^{AB}{}_{cd}V_{AB}{}^{ab} = \frac{1}{2}(\delta^{a}_{c}\delta^{b}_{d} - \delta^{a}_{d}\delta^{b}_{c}) - \frac{1}{8}\Omega^{ab}\Omega_{cd}
\eea
The vielbein also satisfies the following cubic identity descending from the underlying $E_6$ structure:
\begin{equation}
\begin{split}
 & (V_{AB})^{a}{}_{b}(V_{CD})^{b}{}_{c}(V_{EF})^{c}{}_{a} = \frac{1}{2}(\Omega_{A[C}\Omega_{D][E}\Omega_{F]B} - \Omega_{B[C}\Omega_{D][E}\Omega_{F]A})+\\ &+ \frac{1}{16}\Omega_{AB}(\Omega_{C[E}\Omega_{F]D} - \Omega_{D[E}\Omega_{F]C})+ \frac{1}{16}\Omega_{CD}(\Omega_{A[E}\Omega_{F]B} - \Omega_{B[E}\Omega_{F]A})\\ &+ \frac{1}{16}\Omega_{EF}(\Omega_{A[C}\Omega_{D]B} - \Omega_{B[C}\Omega_{D]A})+ \frac{1}{32}\Omega_{AB}\Omega_{CD}\Omega_{EF}.
\end{split}
\end{equation}
Note that there is a small typo in this equation as presented in \cite{Gunaydin}. This is worth a note because \cite{Gunaydin} is surprisingly free of typos for a paper of that size.

\section{$SL(6, \mathbb{R}) \times SL(2, \mathbb{R})$ Structure of the $E_{6(6)}$ Algebra}
\label{sl6appendix}
The five dimensional supergravity theory that we are considering here has an underlying gauge group $SO(p,6-p)$. One of the maximal subgroup of $E_{6(6)}$ is $SL(6, \mathbb{R}) \times SL(2, \mathbb{R})$. As $SO(p,6-p)$ is an obvious subgroup of $SL(6, \mathbb{R}) $, it is convenient to write the $E_{6(6)}$ Lie algebra in a $SL(6, \mathbb{R}) \times SL(2, \mathbb{R})$ basis. In this section, we will very closely follow \cite{Gunaydin}, the only reason we repeat these formulas here is because they are indispensable for the calculations in this paper. 

The fundamental representation of $E_{6(6)}$ under the subgroup $SL(6, \mathbb{R}) \times SL(2, \mathbb{R})$ breaks down into 
\bea
{\bf 27} = ({\bf \overline{15}},{\bf 1}) + ({\bf 6},{\bf 2})
\eea
and the adjoint representation decomposes as 
\bea
{\bf78} = ({\bf35}, {\bf1}) + ({\bf1},{\bf3}) + ({\bf20},{\bf 2}).
\eea
In all of the discussion to follow, the Roman indices $I,J, K...$ run from $1$ to $6$ while the Greek letters $\alpha, \beta...$ run from $1$ to $2$. Under the subgroup, the basis $z^{AB}$ splits into
\bea
z^{AB} = \left(z_{IJ}, z^{I \alpha} \right)
\eea
The vielbein and its inverse decomposes as
\bea
{{V_{A B}}^{a b}=\left(V^{I J a b}, {V_{I \alpha}}^{a b}\right)}
\eea
\bea
{\widetilde{V}^{AB}}{}_{ab}=\left(\tilde{V}_{I J a b}, \tilde{V}^{I \alpha}{}_{a b}\right)
\eea
in such a way that the following relation is satisfied
\bea
\begin{aligned}
\widetilde{V}_{c d}{}^{A B}  V_{A B}{}^{a b} &=\widetilde{V}_{c d I J} \  V^{I J a b}+\widetilde{V}_{c d}{}^{I \alpha}  V_{I \alpha}{}^{a b} \\
&=\delta_{c d}^{a b}+\frac{1}{8} \Omega^{a b}  \Omega_{c d}.
\end{aligned}
\eea
This immediately gives us the identities
\bea
\widetilde{V}_{a b K L} V^{I J a b} &=\delta_{K L}^{I J}
\eea
\bea
\widetilde{V}_{a b}{}^{I \alpha} V_{J \beta}{}^{a b} &=\delta_{J}^{I} \delta_{\beta}^{\alpha}
\eea
\bea
\widetilde{V}_{a b}{}^{I \alpha} V_{K L}{}^{ab} &=\widetilde{V}_{a b K L} V^{I \alpha a b}=0
\eea
Now let us shift focus to the decomposition of the Lie algebra under the subgroup. An infinitesimal action of the full $E_{6(6)}$ group can be realized on $ \left(z_{IJ}, z^{I \alpha} \right)$ as
\bea
\delta z^{A B}=\left(\begin{array}{c}
{\delta z_{I J}} \\
{\delta z^{K \alpha}}
\end{array}\right)=\left(\begin{array}{cc}
{-4 \Lambda^{[M}{}_{[I} \delta^{\left.N\right]} _{J]}} & {\sqrt{2} \ \Sigma_{I J P \beta}} \\
{\sqrt{2} \  \Sigma^{M N K \alpha}} & {\Lambda^{K}{}_{P} \delta_{\beta}^{\alpha}+\Lambda^{\alpha}{}_{\beta} \delta_{P}^{K}}
\end{array}\right)\left(\begin{array}{c}
{z_{M N}} \\
{z^{P \beta}}
\end{array}\right) \label{inftrans}
\eea
where 
\bea
\Sigma_{I J K \alpha}=\Sigma_{[I J K] \alpha}=\frac{1}{6} \epsilon_{I J K M N P} \epsilon_{\alpha \beta} \Sigma^{M N P \beta}, \quad \Lambda^{I}{}_{I}=\Lambda^{\alpha}{}_{\alpha}=0
\eea
All the generators are real and $ \epsilon_{I J K M N P}$, $ \epsilon_{\alpha \beta}$ are the Levi-Civita tensors in $6$ and $4$ dimensions respectively. The equation \eqref{inftrans} can be written more compactly as
\bea\left(\begin{array}{c}
{\delta z_{I J}} \\
{\delta z^{K \alpha}}
\end{array}\right)= \hat{X}\left(\begin{array}{c}
{z_{M N}} \\
{z^{P \beta}}
\end{array}\right) \label{xmatrix}
\eea

Now let us provide a prescription to translate between the $SL(6, \mathbb{R}) \times SL(2, \mathbb{R})$ and the $USp(8)$ basis. This can be done with the help of seven \textit{antisymmetric}, Hermitian $SO(7)$ gamma matrices $\Gamma_i$  satisfying the relations
\bea
{\left\{\Gamma_{i}, \Gamma_{j}\right\}=2 \delta_{i j}}
\eea
\bea
{\Gamma_{0} \Gamma_{1} \ldots \Gamma_{6}=-i \mathbbm{1}}
\eea
We can choose to identify the raising operator $\Omega^{ab}$ as
\bea
\Omega^{ab} = -i \left(\Gamma_{0}\right)^{ab} = -\Omega_{ab}
\eea
Let us define the following quantities
\bea
\Gamma_{I \alpha} =  \left(\Gamma_{I},i \Gamma_{I}\Gamma_{0} \right) \ \ \ \text{and} \ \ \  \Gamma_{I J} = \left[\Gamma_{i}, \Gamma_{j}\right]
\eea
where $\alpha = 1,2$. This lets us decompose $z^{AB}$ and $\tilde{z}^{AB}$ into the $SL(6, \mathbb{R}) \times SL(2, \mathbb{R})$ basis as 
\bea
z^{A B}=\frac{1}{4}\left(\Gamma_{I J}\right)^{A B} z_{I J}+\frac{1}{2} \sqrt{\frac{1}{2}}\left(\Gamma_{I \alpha}\right)^{A B} z^{I \alpha} \label{spliteq1}
\eea
\bea
\tilde{z}_{A B}=\frac{1}{4}\left(\Gamma_{I J}\right)^{A B} \tilde{z}^{I J}-\frac{1}{2} \sqrt{\frac{1}{2}}\left(\Gamma_{I \alpha}\right)^{A B} \tilde{z}_{I \alpha}. \label{spliteq2}
\eea
This split guarantees that the following relation is satisfied
\bea
\tilde{z}_{A B} z^{A B} =  \tilde{z}^{I J}z_{I J} +\tilde{z}_{I \alpha}z^{I \alpha}.
\eea
Using the ``orthogonality'' of the matrices $\Gamma_{I J}$ and $\Gamma_{I \alpha}$, we can invert the above relations :
\bea
z_{I J} = \frac{-1}{4}\left(\Gamma_{I J}\right)^{A B} z^{AB}
\eea
\bea
z^{I \alpha} = \frac{-1}{\sqrt{8}}\left(\Gamma_{I \alpha}\right)^{A B} z^{AB}
\eea
where the $AB$ indices are being summed over.

\section{An Explicit $SL(6, \mathbb{R}) \times SL(2, \mathbb{R})$ Basis}
\label{explicitbasisapp}
From Appendix \ref{sl6appendix}, we can see that in the $SL(6, \mathbb{R}) \times SL(2, \mathbb{R})$ basis, the Lie algebra of $E_{6(6)}$ is spanned by three generators, ${\Lambda^{I}}_{J}$, ${\Lambda^{\alpha}}_{\beta}$ and $\Sigma_{IJK \alpha}$. 

${\Lambda^{I}}_{J}$ and ${\Lambda^{\alpha}}_{\beta}$ are the generators of $SL(6, \mathbb{R})$ and $SL(2, \mathbb{R})$ Lie algebra. Therefore, they can be represented by real traceless matrices. $\Sigma_{IJK \alpha}$ transform in the $(\textbf{20},\textbf{2})$ of $SL(6, \mathbb{R}) \times SL(2, \mathbb{R})$ and is completely antisymmetric in the $IJK$ indices.

Now let us construct a basis for these generators. The Lie algebra of $SL(n, \mathbb{R})$ has dimensions $n^{2}-1$. Therefore, in the fundamental representation, the basis for the $SL(6, \mathbb{R})$ generators would consist of $35$ linearly independent real traceless $6 \times 6$ matrices. To construct such a basis, let us define a tensor
\bea
{\left(\tilde{\Lambda}_{IJ}\right)^{K}}_{L} = \delta^{K}_{I}\delta^{L}_{J}
\eea
Subtracting out the trace, we get
\bea
{\left(\Lambda_{IJ}\right)^{K}}_{L} = \delta^{K}_{I}\delta^{L}_{J} -\frac{1}{6}\delta^{K}_{L}\delta^{I}_{J} \label{lambdaIJgen}
\eea
Let us interpret the the indices $IJ$ and $KL$ as \textit{generator} and \textit{matrix} indices respectively. We can immediately see  that the number of independent generators is $\left(6\times 6\right) - 1 = 35$. For every value of the index $IJ$, the generator ${\left(\Lambda_{IJ}\right)^{K}}_{L}$ corresponds to a real traceless $6 \times 6$ matrix. Therefore, ${\left(\Lambda_{IJ}\right)^{K}}_{L}$ qualifies as a basis for the $SL(6, \mathbb{R})$ Lie algebra.

Similarly, one can construct a basis for the $SL(2, \mathbb{R})$ generators as 
\bea
{\left(\Lambda_{\alpha \beta}\right)^{\gamma}}_{\sigma} = \delta^{\gamma}_{\alpha }\delta^{\sigma}_{\beta
} - \frac{1}{2}\delta^{\gamma}_{\sigma}\delta^{\alpha }_{\beta}
\eea

This gives us $35+ 3= 38 $ linearly independent basis elements. We know that the $E_{6(6)}$ Lie algebra is spanned by $78$ generators. Therefore, we should construct a basis for the $\Sigma_{IJK \alpha}$ generators consisting of $40$ elements. We start off by considering the tensor 
\bea
(\tilde{\Sigma}_{abcd})_{efgh} = \delta_{[a}^{\left[e\right.} \delta_{b}^{g} \delta_{c}^{f}\delta_{d]}^{h]},
\eea
where all the indices run from $1$ to $8$. Now let us restrict the indices $d$ and $h$ to $(7,8)$ and all other indices to (1,...,6). Identifying these indices with $I,J,K,...$ and $\alpha, \beta ,...$ we get the required basis $(\Sigma_{IJK\alpha})_{PQR \beta}$. As the generator is totally antisymmetric in $IJK$ indices, the number of linearly independent elements is $\left(\frac{6 \times 5 \times 4}{3 \times 2}\right) \times 2= 40$.

We will suppress the generator indices from now on to avoid any ambiguity, unless specified otherwise. It was shown in \cite{Gunaydin} that the non-compact part of $E_{6(6)}$ is generated by the symmetric part of the ${\Lambda^{I}}_{J}$ and ${\Lambda^{\alpha}}_{\beta}$ and the self-dual part of the $\Sigma_{IJK \alpha}$. These matrices generate the non-compact coset $E_{6(6)}/USp(8)$ and therefore will be immediately relevant to the calculation of the scalar potential. 


The basis for the symmetric part of $\Lambda_{J}^{I}$ is given by
\bea
{\left(S_{IJ}\right)^{K}}_{L} = \delta^{K}_{(I}\delta^{L}_{J)} -\frac{1}{6}\delta^{K}_{L}\delta^{I}_{J}.
\eea
Similarly, we obtain a basis for the symmetric part of $\Lambda_{\alpha}^{\beta}$:
\bea
{\left(S_{\alpha \beta}\right)^{\gamma}}_{\sigma} =  \delta^{\gamma}_{(\alpha }\delta^{\sigma}_{\beta)
} - \frac{1}{2}\delta^{\gamma}_{\sigma}\delta^{\alpha }_{\beta}.
\eea
The rest of the basis elements can be constructed by taking the self dual part of $(\Sigma_{IJK\alpha})_{PQR \beta}$ with respect to both the generator and the matrix indices. The self dual part of  $(\Sigma_{PQR \beta})_{IJK \alpha}$ w.r.t the matrix indices, denoted by $(\tilde{\Sigma}_{PQR \beta})_{I J K \alpha}^{+}$, satisfies the condition \cite{Gunaydin}
\bea
(\tilde{\Sigma}^{+}_{PQR \beta})_{I J K \alpha}=+ \frac{1}{6} \varepsilon_{\alpha \gamma} \varepsilon_{I J K L M N} (\tilde{\Sigma}^{+}_{PQR \beta})_{L M N \gamma}
\eea
Therefore, the self-dual part of the tensor (w.r.t the matrix indices) will take the form
\bea
(\tilde{\Sigma}^{+}_{PQR \beta})_{I J K \alpha}=(\Sigma_{PQR \beta})_{IJK \alpha}- \frac{1}{6} \varepsilon_{\alpha \gamma} \varepsilon_{I J K L M N} (\Sigma_{PQR \beta})_{L M N \gamma}
\eea
We can do a similar transformation w.r.t the generator indices of $(\tilde{\Sigma}^{+}_{PQR \beta})_{I J K \alpha}$. This gives us the required generator $(\Sigma^{+}_{PQR \beta})_{I J K \alpha}$. 

As ${S^{K}}_{L}$ and ${S^{\alpha}}_{\beta}$ are symmetric and traceless, there will be $\left(\frac{6 \times 7}{2}\right) -1 =20$ and $\left(\frac{2 \times 3}{2}\right) -1 =2$ linearly independent matrices. In the case of $\Sigma^{+}_{I J K \alpha}$, there will be $40/2 =20$ basis elements. This gives us a total of 42 generators and it coincides with the number of non-compact generators of $E_{6(6)}$, as expected.

\section{A Useful Set of $SO(7)$ Gamma matrices}

\noindent
We have already seen that the $SO(7)$ gamma matrices are used to translate between the $USp(8)$ and the $SL(6,\mathbb{R}) \times SL(2,\mathbb{R})$ basis. In this section, we give an explicit construction of the $SO(7)$ gamma matrices that we use in our code. A useful discussion of related ideas can be found in \cite{Murayama}.

The $SO(7)$ gamma matrices are a set of seven Hermitian skew-symmetric matrices $\Gamma_{i}$, with $i = 0,1,...,6$. These satisfy the Clifford algebra:
\bea
\{\Gamma_{i},\Gamma_{j}\} = 2\delta_{ij}
\eea
We also have the following two identities:
\bea
\Gamma_{0}\Gamma_{1}...\Gamma_{6} = -i \mathbbm{1}, \ \ 
\Omega^{ab} = -i (\Gamma_{0})^{ab} = -\Omega_{ab}
\eea

In $d$ dimensions, a representation of the Clifford algebra is constructed from a tensor product of $\lfloor d/2 \rfloor$ Pauli matrices. Thus, the gamma matrices for $SO(7)$ are $8 \times 8$ matrices. We start off with the following naively defined gamma matrices:
\bea
\gamma_1 = \sigma_1 \otimes \mathbbm{1} \otimes \mathbbm{1} \\
\gamma_2 = \sigma_2 \otimes \mathbbm{1} \otimes \mathbbm{1} \\
\gamma_3 = \sigma_3 \otimes \sigma_1 \otimes \mathbbm{1} \\
\gamma_4 = \sigma_3 \otimes \sigma_2 \otimes\mathbbm{1} \\
\gamma_5 = \sigma_3 \otimes \sigma_3 \otimes \sigma_1 \\
\gamma_6 = \sigma_3 \otimes \sigma_3 \otimes \sigma_2 \\
\gamma_0 = \sigma_3 \otimes \sigma_3 \otimes \sigma_3 
\eea

However these are not the gamma matrices that we will use, because these are not Hermitian. We will use some slight modifications to define the gamma matrices for translating between the $USp(8)$ and $SL(6,\mathbb{R}) \times SL(2,\mathbb{R})$ bases. 

First we consider the charge conjugation matrix, which is written as:
\bea
\mathcal{C} = \sigma_1 \otimes i \sigma_2 \otimes \sigma_1
\eea
Under charge conjugation, our naive gamma matrices transform as:
\bea
\mathcal{C}\gamma^i \mathcal{C}^{-1} = -\gamma^{iT}
\eea
Starting from this relation, we look to translate the gamma matrices to a basis where they are all antisymmetric matrices. We diagonalize the charge conjugation matrix, and then re-scale this diagonal matrix to the identity. This is achieved using a unitary matrix $U$:
\bea
U\mathcal{C}U^{T} = \mathbbm{1}
\eea
\noindent
$U$ is explicitly given by:
\bea
U = \frac{1}{\sqrt{2}} \exp{\left[\text{diag}\left(\frac{3 i\pi}{4},\frac{3i\pi}{4},\frac{3i\pi}{4},\frac{3i\pi}{4},\frac{i\pi}{4},\frac{i\pi}{4},\frac{i\pi}{4},\frac{i\pi}{4}\right)\right]} B
\eea
Here, $B$ is the matrix composed of the eigenvectors of $\mathcal{C}$.

Now, to define a set of antisymmetric Hermitian $SO(7)$ gamma matrices, we use:
\bea
\Gamma^i = (U^{T})^{-1} \gamma^{i} U^{T} \quad ; \quad i = 0,1,...6
\eea

\noindent
These seven matrices $\Gamma^i$ are the matrices we use to translate between the $USp(8)$ and $SL(6,\mathbb{R}) \times SL(2,\mathbb{R})$ bases. They are antisymmetric and Hermitian matrices, as required.

\section{Exponentiating the Generators}
\label{exponentiatingappendix}
A crucial step in the calculation of the potential is the construction of the $27$-bein. To do this, first we start off with the $\hat{X}$ matrix defined in \eqref{xmatrix}. This matrix encodes the infinitesimal action of $E_6$ on $(z_{IJ},z^{I\alpha})$ and therefore to find the finite action of the group, we have to exponentiate this matrix appropriately. 

From \eqref{inftrans}, we can see that $\hat{X}$ has four blocks and we can group the indices on each of these blocks into two pairs - an antisymmetrized $IJ$ and $K \alpha$. As $I$, $J$ and $K$ run from $1$ to $6$ and $\alpha$ runs from $1$ to $2$, these pairs will have $15$ and $12$ independent components respectively. In terms of these independent components, the X matrix will have the following block structure:
\bea
\left(\begin{array}{cc}
{15 \times 15} & {15 \times 12} \\
{12 \times 15} & {12 \times 12}
\end{array}\right) \label{eqblock}
\eea

Stacking these blocks upon each other gives us a $27 \times 27$ matrix. This is what we are going to exponentiate to obtain the finite action of the group $E_6$.

To identify the independent components of each of the pairs of indices, we define two sets of bases - one for all the $6 \times 6$ antisymmetric matrices and one for all the $6 \times 2$ matrices. Following the philosophy of Appendix \ref{explicitbasisapp}, we can construct $\mathcal{A}_{aIJ}$ and $\mathcal{B}_{iI \alpha} $, where $a$ and $i$ run from $1$ to $15$ and $1$ to $12$ respectively.

This allows us to define 
\bea
U = \exp{(\hat{X})}.
\eea
Breaking down the matrix $U$ into the form in \eqref{eqblock} and using the bases $\mathcal{A}_{aIJ}$ and $\mathcal{B}_{iI \alpha} $ to translate back to the $I,J,K$ and $\alpha$ indices, we get 
\bea
U = \left(\begin{array}{cc}
{ U^{M N}{}_{I J}} & {U_{P \beta I J}} \\
{U^{I J K \alpha}} & {U_{P \beta}{}^{K \alpha}}
\end{array}\right) 
\eea
In terms of these blocks, we can write the finite action of $E_6$ on $(z_{IJ},z^{I\alpha})$ as
\bea
\begin{aligned}
z_{I J}^{\prime} &=\frac{1}{2} U^{M N}{}_{I J} z_{M N}+U_{P \beta I J} z^{P \beta} \\
z^{\prime K \alpha} &=U_{P \beta}{}^{K \alpha}z ^{P \beta}+\frac{1}{2} U^{I J K \alpha}z_{I J} \label{eqfulltrans}
\end{aligned}
\eea
Combining \eqref{eqfulltrans} with \eqref{spliteq1} and \eqref{spliteq2}, and comparing it with \eqref{vielbeineq}, we get 
\bea
{V^{I J a b}=\frac{1}{8}\left[\left(\Gamma_{K L}\right)^{a b} U_{K L}{}^{I J}+2\left(\Gamma_{K \beta}\right)^{a b} U^{I J K \beta}\right]}
\eea
\bea
{V_{I \alpha}{}^{a b}=\frac{1}{4} \sqrt{\frac{1}{2}}\left[\left(\Gamma_{K L}\right)^{a b} U_{I \alpha K L}+2\left(\Gamma_{K \beta}\right)^{a b} U_{I \alpha}{}^{K \beta}\right]}
\eea
Using the above quantities and \eqref{spliteq1}-\eqref{spliteq2}, we can calculate the $27$-bein in the $USp(8)$ basis as
\bea
V_{A B}{}^{ab}=\frac{1}{4}\left(\Gamma_{I J}\right)^{A B} V^{I J ab}-\frac{1}{2} \sqrt{\frac{1}{2}}\left(\Gamma_{I \alpha}\right)^{A B} V_{I \alpha}{}^{ab}. \label{vielbeineq}
\eea

\section{The Loss Functions}
\label{lossfunctionapp}
\subsection{Gradient Squared Loss Function}
The critical point of any potential is characterized by the vanishing of its gradient. This can be easily implemented in TensorFlow using the command \code{tf.gradients()}. With the help of a normal distribution, we randomly pick a $42$ dimensional array and calculate the value of the potential $P$ at this point on the scalar manifold. The \code{tf.gradients()} command will return the gradient of the potential $\partial_{i}P$ as a $42$ dimensional array. This lets us to define the loss function as:
\bea
\mathcal{S} = \sum_{i =1}^{42}\left|\partial_{i}P\right|^{2}
\eea
where $i$ indexes every direction on the scalar manifold. Note that we are using ordinary partial derivatives along the various scalar directions and then taking an ordinary mod-squared. The sigma model metric plays no role here because the purpose of this loss function is  merely to set all partial derivatives to zero (ideally). Let us also note one subtlety. The default matrix exponentiation command, \code{tf.linalg.expm()}, is not compatible with this loss function as it restricts the backpropagation through code. This can be rectified by defining a new exponential function, we will use the one used by \cite{Fischbacher}.

\subsection{Q-tensor Loss Function}
 The most straightforward approach in calculating the critical points of the potential is to use the gradient squared loss function described in the previous section. However, this loss function is computationally expensive as it involves $42$ gradient calculations at each iteration. This issue can be handled by doing parts of the gradient calculation analytically. This gives us a new loss function, which we will call the $Q$-tensor loss function.

Our $Q$-tensor loss function is motivated by some similar calculations in the $D=4$ case \cite{torsionpaper}. Since the potential term in our lagrangian \eqref{lagrangian} is
\bea
P=-g^{2}\left[\frac{6}{(45)^{2}}\left(T_{a b}\right)^{2}-\frac{1}{96}\left(A_{a b c d}\right)^{2}\right],
\eea
let us consider the variation of the vielbein of the form
\bea
\tilde{V}^{a b A B} \delta V_{A B}{}^{c d} \equiv \Theta^{a b c d}=\Theta^{[a b c d] |}
\eea
The form of $\Theta^{a b cd}$ ensures that the coset structure of the theory is preserved. One can show that the $A$ and $T$ tensors vary as
\bea
{\delta T_{a b}=\frac{5}{2} \Theta_{(a}{}^{c e f} A_{b) c e f}} 
\eea
\bea
{\delta A_{a b c d}=3 \Theta^{e f}{}_{a[b} W_{c d]| e f}+3 \Theta^{e f}{}_{[b c} W_{d]| a e f}}
\eea
It has been shown in \cite{Gunaydin} that the variation of the potential takes the form
\bea
\delta P = -\frac{1}{4} g^{2} \Theta^{[a b cd]|}\widetilde{Q}_{a b cd}
\eea
where 
\bea
\widetilde{Q}_{a b cd} = \frac{1}{27} A_{e a b c} T^{e}{}_{d}+A_{d a}{}^{f g}\left(\frac{1}{2} A_{b c f g}-\frac{1}{45} \Omega_{b f} T_{c g}\right).
\eea
We can see that this variation vanishes when the antisymmetric and symplectic traceless part of $\widetilde{Q}_{a b cd}$ vanishes, owing to the index structure of $\Theta^{[a b cd]|}$. Therefore, at a critical point, the following tensor should vanish
\bea
Q_{a b cd} \equiv \widetilde{Q}_{[a b cd]|}= \mathcal{Q}_{a b cd}+\frac{3}{2} \Omega_{[ab} \mathcal{Q}_{cd] ef} \Omega^{ef}+\frac{1}{8} \Omega_{[ab} \Omega_{cd]} \Omega^{ef} \Omega^{gh}\mathcal{Q}_{efgh} 
\eea
where $\mathcal{Q}_{a b cd} \equiv \widetilde{Q}_{[a b cd]}$ (note the presence and/or of absence of symplectic traces in some of  these definitions). This lets us define the loss function as 
\bea
\mathcal{S} = \sum_{i,j,k,l}|Q_{ijkl}|^{2} \label{qloss}
\eea
\subsection{SUSY Loss Function}
Loss functions have the property that they can be designed to search for certain classes of critical points. One such important class is the one where there are unbroken SUSY generators at the critical point. Such a loss function can be easily engineered by using the fact that there exists at least one massless gravitino at a supersymmetric critical point. Operationally, it amounts to adding a term to \eqref{qloss} which is zero only when there is a massless gravitino. As gravitino masses are the eigenvalues of the gravitino mass matrix $M_{\psi}$, there should exist a vector $\eta$ such that 
\bea
L_{S}:=\left|\left(M_{\psi}\right)^{J}{}_{K} \eta^{K}- \eta^{J}\right|^{2}
\eea
vanishes. Since we were are working with $42$ scalars, we can use the $SO(6)$ symmetry to choose a specific form for $\eta$. In particular, it suffices to pick $\eta^{J} = \delta^{J}_{0}$. Therefore, the full loss function is 
\bea
\mathcal{S}_{SUSY} = \mathcal{S} + \lambda L_{S}
\eea
We have introduced a weight factor $\lambda$ to the SUSY part of the loss function. Choosing $\lambda \sim 10$ and `BFGS' algorithm to find the critical points, we have managed to obtain only $2$ supersymmetric critical points, both of which were previously known. 


\section{Reducing the Parameters}
The coset manifold is spanned by three sets of generators in the $SL(6, \mathbb{R}) \times SL(2, \mathbb{R})$ basis - ${S^{I}}_{J}$, ${S^{\alpha}}_{\beta}$ and $\Sigma^{+}_{IJK \alpha}$. This space is parametrized by $42$ scalars. Therefore, finding a critical point of the potential amounts to solving a $42$ parameter loss function optimization problem in TensorFlow. The potential is invariant under the action of the group $SO(6) \times SU(1,1)$. We can make use of this symmetry to reduce the number of parameters of the optimization problem. 

Consider the first set of generators. Reinstating the generator indices, we can see that they have the form ${\left(S_{IJ}\right)^{K}}_{L}$ and are $20$ in number. (Refer Appendix \ref{explicitbasisapp} for more details on the construction of the generator). The generator indices $I$ and $J$ run from $1$ to $6$. The $SO(6)$ group acts on these indices through its fundamental representation. As these group elements are the rotation matrices, we can use them to diagonalize the generator indices of ${\left(S_{IJ}\right)^{K}}_{L}$. This leaves us with $20 -\frac{6 \times 5}{2} =5$ basis elements.

Now let use the remaining $SU(1,1)$ symmetry to reduce more parameters. As $SL(2, \mathbb{R}) \cong SU(1,1)$, we can clearly see that the two scalars corresponding to ${S^{\alpha}}_{\beta}$ are redundant and does not contribute to the calculation of the potential.

There are only two $SU(3)$ singlets that lie outside the $SL(2, \mathbb{R})$ and these singlets rotate into each other under the action of the $U(1)$ subgroup of $SU(1,1)$. Both the singlets are generated by the basis elements
$(\Sigma^{+}_{PQR \beta})_{I J K \alpha}$. Therefore, the scalars corresponding to these singlets contain redundant information. To take care of this, let us focus on one of the singlets generated by 
\bea
-\Sigma^{+}_{1351}+\Sigma^{+}_{1362}+\Sigma^{+}_{1452}+\Sigma^{+}_{1461}
\eea
where we have suppressed the \textit{matrix} indices for brevity. First, we perform a basis change from 
\bea
\begin{aligned}
 \left(\Sigma^{+}_{1351},\Sigma^{+}_{1362},\Sigma^{+}_{1452},\Sigma^{+}_{1461}\right) \to \hspace{1.2in} \\ \Bigl{(} -\Sigma^{+}_{1351}+\Sigma^{+}_{1362}+\Sigma^{+}_{1452}+\Sigma^{+}_{1461}, 
\Sigma^{+}_{1351}-\Sigma^{+}_{1362}+\Sigma^{+}_{1452}+\Sigma^{+}_{1461},  \\
\Sigma^{+}_{1351}+\Sigma^{+}_{1362}-\Sigma^{+}_{1452}+\Sigma^{+}_{1461},
\Sigma^{+}_{1351}+\Sigma^{+}_{1362}+\Sigma^{+}_{1452}-\Sigma^{+}_{1461}\Bigl{)}
\end{aligned}
\eea
We can easily see that the scalar corresponding to the first transformed basis element is redundant. Therefore, we can just drop it from the construction of the potential. This leaves us with $20-1 =19$ $\Sigma^{+}_{IJK \alpha}$ generators.

We have managed to reduce the number of parameters of the problem from $42$ to $24$ by utilizing the symmetry of the potential. It turns out that this reduced parameter approach is the most efficient and quickest way to get all the critical points we have reported in this paper.

\section{Critical Points}
\label{criticalpointappendix}

We now list the critical points, with their potential values, gravitino masses $m^{2}/m_{0}^{2}(\psi)$, supersymmetry and the location on the scalar manifold. For critical points with a residual symmetry, there is a continuum of points on the scalar manifold corresponding to the critical point (and also there can be discrete redundancies). We only give one representative set of such scalar values.

The location of the critical points are labeled by the values of the three sets of scalars, $\phi_{IJ},\phi_{IJK \alpha}$ and $\phi_{\alpha \beta} $, having $20$, $20$ and $2$ independent components respectively. We will report these independent components as the elements of three arrays :
\bea
   	\begin{matrix}
\phi_{IJ} =[&\phi_{11} & \phi_{12} & \phi_{13} & \phi_{14} &\phi_{15}&\\
&\phi_{16} & \phi_{22} & \phi_{23} &\phi_{24} &\phi_{25}&\\
  &\phi_{26} & \phi_{33} & \phi_{34} &\phi_{35} &\phi_{36}&\\
 &\phi_{44} & \phi_{45} & \phi_{46} &\phi_{55} &\phi_{56} &   ]
\end{matrix}
\eea
\bea
   	\begin{matrix}
\phi_{IJK \alpha} =[&\phi_{0 1  2  0} & \phi_{ 0 1 3 1} & \phi_{0 1 4 0 } & \phi_{0 1 5 1 } &\phi_{0 2 3 0}&\\
&\phi_{0 2 4 1} & \phi_{ 0 2 5 0} & \phi_{0 3 4 0} &\phi_{0 3 5 1 } &\phi_{0 4 5 0}&\\
  &\phi_{0 1 2 1} & \phi_{0 1 3 0} & \phi_{0 1 4 1} &\phi_{0 1 5 0} &\phi_{0 2 3 1}&\\
 &\phi_{0 2 4 0} & \phi_{0 2 5 1} & \phi_{0 3  4 1} &\phi_{0 3 5 0} &\phi_{0 4 5 1} &   ]
\end{matrix}
\eea
\bea
   	\begin{matrix}
\phi_{\alpha \beta} =[&\phi_{11} &\phi_{12} &   ]
\end{matrix}
\eea
The gravitino masses $m^{2}/m_{0}^{2}(\psi)$ will be reported as $m_{n}$, where $m$ is the mass of the gravitino and $n$ is the multiplicity of the mass.
 \begin{enumerate}
     \item Potential = -0.75$g^2$ \newline
           $m^{2}/m_{0}^{2}(\psi) = 1_8$ \newline
           SUSY = 8 \newline
            $\phi_{IJK \alpha} : 0_{20}$ \newline
           $\phi_{IJ} \phantom{ab}: 0_{20}$ \newline
             $\phi_{\alpha \beta} \phantom{ab}: 0_{2}$
\vspace{0.1in}
\newline
\vspace{0.1in}
 \noindent\rule{15cm}{0.4pt}
     \item Potential = -0.78003$g^2$ \newline
           $m^{2}/m_{0}^{2}(\psi) = 1.1852_8$ \newline
           SUSY = 0 \newline 
$\phi_{IJK \alpha}:  \phantom{a}0_{20}$

\noindent$\begin{matrix}
\phi_{IJ} \phantom{ab} : [& \phantom{-}0.0220  & \phantom{-}0.0721  & \phantom{-}0.0689  & \phantom{-}0.0437
  & \phantom{-}0.2364&\\
&-0.2048 &-0.0805  & \phantom{-}0.0217
  & \phantom{-}0.0138  & \phantom{-}0.0747&\\
  &-0.0647 &-0.0815
  & \phantom{-}0.0132  & \phantom{-}0.0714 &-0.0618&\\
 &-0.0877
  & \phantom{-}0.0453 &-0.0392  & \phantom{-}0.0306 &-0.2123& \phantom{-}  ]
\end{matrix}$

\noindent$   	\begin{matrix}
\phi_{\alpha \beta} \phantom{ab} : [&-0.0033 &-0.0129 & \phantom{-}  ]
\end{matrix}$
\vspace{0.1in}
\newline
\vspace{0.1in}
 \noindent\rule{15cm}{0.4pt}
\item Potential = -0.83995$g^2$ \newline
       $m^{2}/m_{0}^{2}(\psi) = 1_2, 1.3611_4, 1.7778_2$ \newline
       SUSY = $2$

\noindent$\begin{matrix}
\phi_{IJK \alpha} : \hspace{0.1in} \bigl{[}& \phantom{-}0.2054
& \phantom{-}0.2253
& -0.3615
& -0.0837
& -0.0763
&\\
& \phantom{-}0.0171
& -0.3377
& \phantom{-}0.2924
& \phantom{-}0.1268
& -0.1356
&\\
& \phantom{-}0.2724
& \phantom{-}0.1625
& \phantom{-}0.26
& -0.0174
& \phantom{-}0.3851
&\\
& -0.2551
& \phantom{-}0.0658
& \phantom{-}0.1204
& -0.3346
& -0.4661
& \phantom{-}\bigl{]}
\end{matrix}$

\noindent $\begin{matrix}
\phi_{IJ}  \phantom{ab} : \hspace{0.1in} \bigl{[}& -0.1873
& -0.0891
& -0.0447
& \phantom{-}0.0679
& \phantom{-}0.0485
&\\
& \phantom{-}0.0032
& -0.1095
& \phantom{-}0.0859
& -0.1501
& -0.0966
&\\
& \phantom{-}0.0466
& -0.1773
& -0.0659
& -0.062
& -0.0788
&\\
& -0.1511
& \phantom{-}0.0739
& -0.0332
& -0.1763
& \phantom{-}0.0588
& \phantom{-} \bigl{]}
\end{matrix} $

\noindent$	\begin{matrix}
\phi_{\alpha \beta} \phantom{ab} : \hspace{0.1in} \bigl{[}&-0.1311 & \phantom{-}
0.0201 & \phantom{-}   \bigl{]}
\end{matrix}$
\vspace{0.1in}
\newline
 \noindent\rule{15cm}{0.4pt}
\item Potential = -0.84375$g^2$ \newline
      $m^{2}/m_{0}^{2}(\psi) = 1.2099_6, 2_2$ \newline
      SUSY = 0

\noindent$   	\begin{matrix}
\phi_{IJK \alpha} :  \bigl{[}& -0.2121
& \phantom{-}0.4605
& -0.4034
& -0.1082
& \phantom{-}0.4559
&\\
& \phantom{-}0.2166
& -0.3608
& -0.0295
& \phantom{-}0.3122
& \phantom{-}0.0428
&\\
& -0.4367
& -0.3318
& -0.0640
& -0.3174
& \phantom{-}0.1562
&\\
& -0.1349
& \phantom{-}0.3731
& \phantom{-}0.2580
& \phantom{-}0.2858
& \phantom{-}0.2570 & \phantom{-}   \bigl{]}
\end{matrix}$ \newline
$
\phi_{IJ}  \phantom{ab} : \phantom{a}0_{20}
$

\noindent $   	\begin{matrix}
\phi_{\alpha \beta} \phantom{ab} : \bigl{[}& \phantom{-}0.1736 & \phantom{-}0.0409& \phantom{-}   \bigl{]}
\end{matrix}$
\vspace{0.1in}
\newline
\vspace{0.1in}
 \noindent\rule{15cm}{0.4pt}
\item Potential = -0.870298$g^2$ \newline
      $m^{2}/m_{0}^{2}(\psi) = 1.4399_4, 1.6000_4$ \newline
      SUSY = 0

\noindent$   	\begin{matrix}
\phi_{IJK \alpha} :  \bigl{[}& \phantom{-}0.1731  & \phantom{-}0.1489  & \phantom{-}0.0199  & \phantom{-}0.1498 &-0.0209 &\\ & \phantom{-}0.0468
  & \phantom{-}0.1180 &-0.2987 &-0.3286  & \phantom{-}0.1059 &\\ &-0.0215 &-0.5504
  & \phantom{-}0.0340  &-0.0369 &-0.0315&\\ &-0.4731 &-0.1205 &-0.0286
 &-0.1302 &-0.0505 & \phantom{-}   \bigl{]}
\end{matrix}$

\noindent$\begin{matrix}
\phi_{IJ}  \phantom{ab} : \bigl{[}& -0.1354  & \phantom{-}0.2437 &-0.0080  & \phantom{-}0.1156
 &-0.1964 &\\ & \phantom{-}0.0244 &-0.2594 &-0.0288  & \phantom{-}0.0411 &-0.0775
&\\  & \phantom{-}0.1856  &-0.3427  & \phantom{-}0.0119 &-0.0159 &-0.0993&\\ &-0.3271
 &-0.0727 &-0.0931 &-0.2911  & \phantom{-}0.1274& \phantom{-}   \bigl{]}
\end{matrix}$

\noindent $   	\begin{matrix}
\phi_{\alpha \beta} \phantom{ab} : \bigl{[}&-0.1913 &-0.6837& \phantom{-}  \bigl{]}
\end{matrix}$
\vspace{0.1in}
\newline
\vspace{0.1in}
 \noindent\rule{15cm}{0.4pt}
\item Potential = -0.87894$g^2$ \newline
      $m^{2}/m_{0}^{2}(\psi) = 1.3576_4, 1.8020_4$ \newline
      SUSY = 0

\noindent $   	\begin{matrix}
\phi_{IJK \alpha} : \bigl{[}& \phantom{-}0.2806 &-0.0487  & \phantom{-}0.0370  & \phantom{-}0.1231  & \phantom{-}0.0244 &\\  & \phantom{-}0.1272
 &-0.1091 &-0.4679 &-0.2843  & \phantom{-}0.1845&\\ &-0.1222 &-0.1561
 &-0.0124 &-0.4571  & \phantom{-}0.2290&\\ &-0.4114 & \phantom{-}0.0464  & \phantom{-}0.1105
  & \phantom{-}0.0659 &-0.0464& \phantom{-}  \bigl{]}
\end{matrix}$

\noindent $  \begin{matrix}
\phi_{IJ}  \phantom{ab} : \bigl{[}& -0.1378  & \phantom{-}0.0300 &-0.0118 &-0.0975
  & \phantom{-}0.0313&\\  &-0.0659  & \phantom{-}0.0933  & \phantom{-}0.2983 &-0.0088  & \phantom{-}0.2439&\\
 &-0.0875  &-0.0127   & \phantom{-}0.1837  & \phantom{-}0.0903  & \phantom{-}0.1253&\\   & \phantom{-}0.0380
 &-0.0952   & \phantom{-}0.3257  &-0.0834 &-0.1322& \phantom{-}   \bigl{]}
\end{matrix}$

\noindent $   	\begin{matrix}
\phi_{\alpha \beta} \phantom{ab} : \bigl{[}& -0.2207 &-0.1374& \phantom{-}  \bigl{]}
\end{matrix}$
\vspace{0.1in}
\newline
\vspace{0.1in}
 \noindent\rule{15cm}{0.4pt}
\item Potential = -0.88764$g^2$ \newline
      $m^{2}/m_{0}^{2}(\psi) = 1.4834_2, 1.6041_4, 1.8380_2$ \newline
      SUSY = 0

\noindent $     	\begin{matrix}
\phi_{IJK \alpha} :  \bigl{[}& -0.0175  & \phantom{-}0.4429  & \phantom{-}0.3736  & \phantom{-}0.0638
 &-0.0082&\\  & \phantom{-}0.0819  & \phantom{-}0.2291  & \phantom{-}0.2761
  & \phantom{-}0.1609  & \phantom{-}0.1558&\\ &-0.3163  & \phantom{-}0.2231
  & \phantom{-}0.1980  & \phantom{-}0.2262 &-0.2012&\\  & \phantom{-}0.0965
 &-0.2062 &-0.1198  & \phantom{-}0.0234 &-0.1573& \phantom{-}  \bigl{]}
\end{matrix}$

\noindent $  \begin{matrix}
\phi_{IJ}  \phantom{ab} : \bigl{[}& -0.0627  & \phantom{-}0.0528 &-0.3286 &-0.1493
  & \phantom{-}0.2079&\\  & \phantom{-}0.0001 &-0.3431 &-0.0243
  & \phantom{-}0.0112  & \phantom{-}0.0358&\\ &-0.0811 &-0.2810
  & \phantom{-}0.0849 &-0.1001 &-0.0682&\\ &-0.3193
  & \phantom{-}0.0224 &-0.1788 &-0.3001 &-0.2017& \phantom{-}   \bigl{]}
\end{matrix}$

\noindent $  	\begin{matrix}
\phi_{\alpha \beta} \phantom{ab} : \bigl{[}& \phantom{-}0.3857 &-0.0368& \phantom{-}  \bigl{]}
\end{matrix}$
\vspace{0.1in}
\newline
\vspace{0.1in}
 \noindent\rule{15cm}{0.4pt}
\item Potential = -0.89291$g^2$ \newline
      $m^{2}/m_{0}^{2}(\psi) = 1.6667_8$ \newline
      SUSY = 0

\noindent $   	\begin{matrix}
\phi_{IJK \alpha} :  \bigl{[}& -0.0916  & \phantom{-}0.2859 &-0.5150  & \phantom{-}0.2259 &-0.1448&\\  & \phantom{-}0.2369
 &-0.1957  &-0.0105 &-0.0900  & \phantom{-}0.2174&\\  & \phantom{-}0.0687  & \phantom{-}0.2874
 &-0.3466  & \phantom{-}0.1842    & \phantom{-}0.0161&\\  & \phantom{-}0.0122  & \phantom{-}0.1494  & \phantom{-}0.0193
 &-0.1713  & \phantom{-}0.1907& \phantom{-}  \bigl{]}
\end{matrix}$

\noindent $  \begin{matrix}
\phi_{IJ}  \phantom{ab} : \bigl{[}& -0.1063  & \phantom{-}0.0160  & \phantom{-}0.0169 &-0.1382
  & \phantom{-}0.2505&\\ &-0.0097  &-0.1271 &-0.0277 &-0.1733 &-0.0407&\\
 &-0.1353 &-0.1640  & \phantom{-}0.0405  & \phantom{-}0.0719   & \phantom{-}0.0525&\\  & \phantom{-}0.0994 
 &-0.1144  & \phantom{-}0.3696  & \phantom{-}0.2116  & \phantom{-}0.1617& \phantom{-}   \bigl{]}
\end{matrix}$

\noindent $   	\begin{matrix}
\phi_{\alpha \beta} \phantom{ab} : \bigl{[}& \phantom{-}0.1235  & \phantom{-}0.2164 & \phantom{-}  \bigl{]}
\end{matrix}$
\vspace{0.1in}
\newline
\vspace{0.1in}
 \noindent\rule{15cm}{0.4pt}
\item Potential = -0.96395$g^2$ \newline
      $m^{2}/m_{0}^{2}(\psi) = 1.6244_2, 1.7526_2, 1.9237_2, 2.0629_2$ \newline
      SUSY = 0

\noindent $    	\begin{matrix}
\phi_{IJK \alpha} :  \bigl{[}& -0.3846 &-0.3583  & \phantom{-}0.4398 &-0.1921  & \phantom{-}0.5825&\\  & \phantom{-}0.5682
 &-0.1986 &-0.1895& \phantom{-} 0.3256& \phantom{-} 0.0181&\\& \phantom{-} 0.1190& \phantom{-} 0.5485&
 -0.3227& \phantom{-} 0.1301& -0.2848&\\& -0.5394& -0.4508& \phantom{-} 0.1333&
 -0.1853& \phantom{-} 0.0386 & \phantom{-}  \bigl{]}
\end{matrix}$

\noindent $  \begin{matrix}
\phi_{IJ}  \phantom{ab} : \bigl{[}&-0.0957& -0.1056& \phantom{-} 0.2188& -0.1265&\phantom{-}
  0.1334&\\& -0.0700& -0.0747& \phantom{-} 0.2928& -0.0173& -0.1279&\\&
  \phantom{-}0.0050& -0.1213& -0.0582& -0.0072& \phantom{-} 0.0326&\\&  -0.3010&
 -0.0386& -0.1698& -0.2651& \phantom{-} 0.0327 & \phantom{-}  ]
\end{matrix}$

\noindent $  	\begin{matrix}
\phi_{\alpha \beta} \phantom{ab} : \bigl{[}& \phantom{-}0.0817  &-0.0928& \phantom{-}  \bigl{]}
\end{matrix}$
\vspace{0.1in}
\newline
\vspace{0.1in}
 \noindent\rule{15cm}{0.4pt}
\item Potential = -0.96410$g^2$ \newline
      $m^{2}/m_{0}^{2}(\psi) = 1.7013_4, 1.9845_4$ \newline
      SUSY = 0

\noindent $   	\begin{matrix}
\phi_{IJK \alpha} :  \bigl{[}& -0.0942& -0.1853& \phantom{-} 0.3233& -0.3281& \phantom{-} 0.0565&\\& -0.5403&
 -0.4239& -0.4596& -0.0072& -0.4892&\\& \phantom{-} 0.0531& -0.2818& \phantom{-}
  0.1580& -0.5857& \phantom{-}  0.0732& \phantom{-} \\& \phantom{-}0.3125& \phantom{-} 0.5560& -0.1772&
 \phantom{-} 0.1402& -0.4972 & \phantom{-}   \bigl{]}
\end{matrix}$

\noindent $  \begin{matrix}
\phi_{IJ}  \phantom{ab} : \bigl{[}& \phantom{-}0.2592& -0.0041& \phantom{-} 0.0489& \phantom{-} 0.0640&
  \phantom{-} 0.2615&\\& \phantom{-} 0.0347& \phantom{-} 0.1877& -0.0208& -0.3138& -0.0587&\\&  \phantom{-}
  0.0322& \phantom{-}  0.3474& -0.0307& -0.0509& -0.0318&\\& \phantom{-} 0.1081&  \phantom{-}
  0.1454& -0.0495& \phantom{-} 0.0532& \phantom{-} 0.0372& \phantom{-}  ]
\end{matrix}$

\noindent $    	\begin{matrix}
\phi_{\alpha \beta} \phantom{ab} : \bigl{[}&-0.3236   &-0.1053& \phantom{-}  \bigl{]}
\end{matrix}$
\vspace{0.1in}
\newline
\vspace{0.1in}
 \noindent\rule{15cm}{0.4pt}
\item Potential = -0.96453$g^2$ \newline
      $m^{2}/m_{0}^{2}(\psi) = 1.7982_4, 1.8992_4$ \newline
      SUSY = 0

\noindent $  	\begin{matrix}
\phi_{IJK \alpha} :  \bigl{[}&\phantom{-} 0.0111
&\phantom{-} 0.0568
& \phantom{-}0.2713
& -0.1701
& -0.4522
&\\
& \phantom{-}0.0359
&\phantom{-} 0.0121
& -0.289
& -0.0665
& -0.11
&\\
& \phantom{-}0.1809
& -0.1729
&\phantom{-} 0.1091
& \phantom{-}0.0994
&\phantom{-} 0.7285
&\\
& \phantom{-}0.802
& -0.0795
& -0.6458
& \phantom{-}0.0798
& -0.1303& \phantom{-} \bigl{]}
\end{matrix}$

\noindent $  \begin{matrix}
\phi_{IJ}  \phantom{ab} : \bigl{[}&\phantom{-} 0.0442
&\phantom{-}0.0204
& -0.1018
& -0.0907
&\phantom{-} 0.0135
&\\
& -0.0275
& \phantom{-}0.2802
& -0.1701
& 0.2117
& -0.1261
&\\
&\phantom{-} 0.1647
&\phantom{-} 0.2986
&\phantom{-} 0.1598
& -0.2005
& -0.0628
&\\
& \phantom{-}0.2719
& \phantom{-}0.0484
&\phantom{-} 0.0043
&\phantom{-} 0.2103
& -0.054 & \phantom{-}  ]
\end{matrix}$

\noindent $    	\begin{matrix}
\phi_{\alpha \beta} \phantom{ab} : \bigl{[}&
-0.6799 &
\phantom{-}0.2815& \phantom{-}  \bigl{]} 
\end{matrix}$
\vspace{0.1in}
\newline
\vspace{0.1in}
 \noindent\rule{15cm}{0.4pt}
\item Potential = -0.98278$g^2$ \newline
      $m^{2}/m_{0}^{2}(\psi) = 1.6296_4, 2.2222_4$ \newline
      SUSY = 0

\noindent $    	\begin{matrix}
\phi_{IJK \alpha} :  \bigl{[}& -0.4617& -0.1931& \phantom{-} 0.1004& -0.3002& \phantom{-} 0.2972&\\& \phantom{-}    0.4158& \phantom{-}
  0.1752& -0.3312& \phantom{-} 0.1216& \phantom{-} 0.1789&\\& -0.5863& \phantom{-} 0.4183&\phantom{-}
  0.5573& -0.4897& \phantom{-} 0.1108&\\& \phantom{-} 0.2264& \phantom{-} 0.2604& -0.3574& \phantom{-}
  0.4893& \phantom{-} 0.5069& \phantom{-} \bigl{]}
\end{matrix}$

\noindent $  \begin{matrix}
\phi_{IJ}  \phantom{ab} : \bigl{[}& \phantom{-}0.1698& \phantom{-} 0.0405& -0.2112& \phantom{-} 0.1410&
 -0.0524&\\& -0.0343& \phantom{-} 0.0917& \phantom{-} 0.0768& \phantom{-} 0.1164& \phantom{-} 0.0746&\\&
  \phantom{-} 0.2937& -0.1276& \phantom{-}  0.0075& \phantom{-} 0.0552& \phantom{-} 0.0344&\\& -0.0179&  \phantom{-}
  0.2003& -0.1533& \phantom{-} 0.1154&-0.1315& \phantom{-}   \bigl{]}
\end{matrix}$

\noindent $  	\begin{matrix}
\phi_{\alpha \beta} \phantom{ab} : \bigl{[}&-0.1484 &-0.1214& \phantom{-}  \bigl{]} 
\end{matrix}$
\vspace{0.1in}
\newline
\vspace{0.1in}
 \noindent\rule{15cm}{0.4pt}
\item Potential = -1.00148$g^2$ \newline
      $m^{2}/m_{0}^{2}(\psi) = 1.7434_4, 2.1884_4$ \newline
      SUSY = 0

\noindent $   	\begin{matrix}
\phi_{IJK \alpha} :  \bigl{[}& \phantom{-}0.0734& -0.0501& \phantom{-} 0.2818& \phantom{-} 0.0023&
 -0.3375&\\& -0.1750& \phantom{-} 0.2909& -0.1784&
 -0.1179& \phantom{-} 0.2328&\\& -0.5717& \phantom{-} 0.8423&
 \phantom{-} 0.9184& \phantom{-} 0.1358& -0.3331&\\& -0.0586&  \phantom{-}
  0.6736& -0.3227& -0.1372& \phantom{-} 0.4133 & \phantom{-}   \bigl{]}
\end{matrix}$

\noindent $  \begin{matrix}
\phi_{IJ}  \phantom{ab} : \bigl{[}& -0.1791& -0.0070& -0.1408& \phantom{-} 0.1044&  \phantom{-}
  0.2529&\\& \phantom{-} 0.0596& \phantom{-} 0.0035& -0.0189&
  \phantom{-} 0.0000  & \phantom{-}0.0183&\\& \phantom{-} 0.0006& -0.2790&  \phantom{-}
  0.1338& -0.1567& \phantom{-} 0.0645&\\& -0.0252&
 -0.0437& -0.0272& -0.2297& -0.0175 & \phantom{-}   \bigl{]}
\end{matrix}$

\noindent $  \begin{matrix}
\phi_{\alpha \beta} \phantom{ab} : \bigl{[}&-0.0540 & \phantom{-}0.1104& \phantom{-}  \bigl{]} 
\end{matrix}$
\vspace{0.1in}
\newline
\vspace{0.1in}
 \noindent\rule{15cm}{0.4pt}
\item Potential = -1.05469$g^2$ \newline
      $m^{2}/m_{0}^{2}(\psi) = 1.7333_2, 2.3259_6$ \newline
      SUSY = 0

\noindent $  	\begin{matrix}
\phi_{IJK \alpha} :  \bigl{[}& -0.4281
& \phantom{-}0.4205
& -0.6442
& -0.1746
& \phantom{-}0.0484
&\\
& \phantom{-}0.1198
& -0.6066
& -0.3272
& -0.0898
& -0.6883
&\\
& -0.1756
& \phantom{-}0.1134
& -0.3413
& -0.1811
& \phantom{-}0.0975
&\\
& \phantom{-}0.3744
& \phantom{-}0.2832
& -0.3668
& -0.5122
& -0.0741
& \phantom{-} \bigl{]}
\end{matrix}$

\noindent $  \begin{matrix}
\phi_{IJ}  \phantom{ab} : \bigl{[}& \phantom{-}0.0833
& \phantom{-}0.0302
& -0.0707
& \phantom{-}0.1785
& \phantom{-}0.2904
&\\
& \phantom{-}0.0862
& \phantom{-}0.2312
& \phantom{-}0.1426
& \phantom{-}0.4293
& -0.1833
&\\
& -0.045
& \phantom{-}0.4197
& -0.1146
& \phantom{-}0.1314
& -0.0578
&\\
& \phantom{-}0.2373
& \phantom{-}0.0993
& \phantom{-}0.0559
& \phantom{-}0.3571
& \phantom{-}0.1576
& \phantom{-} \bigl{]}
\end{matrix}$

\noindent $  \begin{matrix}
\phi_{\alpha \beta} \phantom{ab} : \bigl{[}& \phantom{-}0.1934 &  \phantom{-}
0.2085 & \phantom{-} ] 
\end{matrix}$
\vspace{0.1in}
\newline
\vspace{0.1in}
 \noindent\rule{15cm}{0.4pt}
\item Potential = -1.07353$g^2$ \newline
      $m^{2}/m_{0}^{2}(\psi) = 1.8190_2, 2.0110_2, 2.4329_2, 2.8060_2$ \newline
      SUSY = 0

\noindent $  	\begin{matrix}
\phi_{IJK \alpha} :  \bigl{[}&-0.2506& \phantom{-} 0.1669& \phantom{-}  0.2442& -0.0534& -0.0400&\\& -0.3939&  \phantom{-}
  0.4725& -0.4666& \phantom{-} 0.2557& \phantom{-} 0.0713&\\& -0.1074& -0.5340&\phantom{-}
  0.1528& -1.0272&  -0.2950&\\& -0.1738& \phantom{-}  0.3371& \phantom{-} 0.4106&
 -0.2600& -0.1758 & \phantom{-}   \bigl{]}
\end{matrix}$

\noindent $  \begin{matrix}
\phi_{IJ}  \phantom{ab} : \bigl{[}& -0.0919& \phantom{-} 0.0282& -0.1423& \phantom{-} 0.1792&
  \phantom{-} 0.3608&\\& -0.1764& \phantom{-} 0.0773& -0.0911& \phantom{-} 0.1521& \phantom{-}  0.1619&\\& \phantom{-}
  0.4051& -0.1765& \phantom{-} 0.0094& -0.2168& -0.0329&\\& \phantom{-} 0.1590&
 -0.1648& -0.2038& \phantom{-} 0.1308& -0.1027& \phantom{-} \bigl{]}
\end{matrix}$

\noindent $  \begin{matrix}
\phi_{\alpha \beta} \phantom{ab} : \bigl{[}& \phantom{-}0.3044& -0.1611& \phantom{-}  \bigl{]}
\end{matrix}$
\vspace{0.1in}
\newline
\vspace{0.1in}
 \noindent\rule{15cm}{0.4pt}
\item Potential = -1.125$g^2$ \newline
      $m^{2}/m_{0}^{2}(\psi) = 2.4444_8$ \newline
      SUSY = 0

\noindent $  	\begin{matrix}
\phi_{IJK \alpha} :  \bigl{[}& \phantom{-}0.1216& \phantom{-}  0.5489&  -0.3592& \phantom{-} 0.3047& \phantom{-} 0.4132&\\& -0.2438& \phantom{-}
  0.4997& \phantom{-} 0.1565& -0.4299& -0.4129&\\&  -0.4785& \phantom{-} 0.3301& \phantom{-}
  0.0159& -0.6865& \phantom{-}  0.5689&\\& -0.1988& -0.1670& \phantom{-}  0.1878&
 -0.1575& \phantom{-} 0.4359 & \phantom{-}   \bigl{]}
\end{matrix}$

\noindent $  \begin{matrix}
\phi_{IJ}  \phantom{ab} : \bigl{[}&-0.4605& -0.2603& -0.1841& \phantom{-} 0.0171&
 -0.1131&\\& \phantom{-} 0.1103& -0.1645& \phantom{-} 0.2359& \phantom{-} 0.0977& \phantom{-} 0.3436&\\&\phantom{-}
  0.0426& -0.3715& -0.3390& \phantom{-} 0.0086& \phantom{-} 0.1074&\\& -0.1506& \phantom{-}
  0.3230& \phantom{-} 0.1392& -0.4254& -0.1013 & \phantom{-} \bigl{]}
\end{matrix}$

\noindent $  \begin{matrix}
\phi_{\alpha \beta} \phantom{ab} : \bigl{[}&-0.3019 &-0.0469& \phantom{-}  \bigl{]}
\end{matrix}$
\vspace{0.1in}
\newline
\vspace{0.1in}
 \noindent\rule{15cm}{0.4pt}
\item Potential = -1.29725$g^2$ \newline
      $m^{2}/m_{0}^{2}(\psi) = 1.9011_2, 2.5684_2, 2.9636_2, 3.7576_2$ \newline
      SUSY = 0

\noindent $  	\begin{matrix}
\phi_{IJK \alpha} :  \bigl{[}& \phantom{-}0.2989& -0.3199& -1.0217& -0.4704& \phantom{-} 0.0110&\\& -0.1770& \phantom{-}
  0.0661& \phantom{-} 0.2357& -0.0161& \phantom{-} 0.7850&\\& \phantom{-} 0.7253& \phantom{-} 0.2029&
 -0.8632& -0.6994& -0.1939&\\& -0.2820& -0.1526& \phantom{-}  0.0638&
 -0.6484& -0.0747 & \phantom{-}   \bigl{]}
\end{matrix}$

\noindent $  \begin{matrix}
\phi_{IJ}  \phantom{ab} : \bigl{[}& \phantom{-} 0.0408& \phantom{-} 0.0791& -0.0928& -0.1345&
 -0.0212&\\& \phantom{-} 0.0384& -0.1104& \phantom{-} 0.2825& \phantom{-} 0.4864& \phantom{-} 0.0939&\\&\phantom{-}
  0.0695& -0.0916& -0.2976& -0.1065& -0.1207&\\& -0.2503& \phantom{-}
  0.1465& \phantom{-} 0.0311& -0.4924& -0.1575& \phantom{-}  \bigl{]}
\end{matrix}$

\noindent $  \begin{matrix}
\phi_{\alpha \beta} \phantom{ab} : \bigl{[}& \phantom{-}0.3801& -0.0305& \phantom{-}  \bigl{]} 
\end{matrix}$
\vspace{0.1in}
\newline
\vspace{0.1in}
 \noindent\rule{15cm}{0.4pt}
\item Potential = -1.30291$g^2$ \newline
      $m^{2}/m_{0}^{2}(\psi) = 1.8684_2, 2.9387_2, 2.9656_2, 3.7147_2$ \newline
      SUSY = 0

\noindent $  \begin{matrix}
\phi_{IJK \alpha} :  \bigl{[}&-0.2212& -0.4424& \phantom{-} 0.0436& \phantom{-} 0.6466& -1.3871&\\& \phantom{-}  0.1276& \phantom{-}
  0.3229& \phantom{-} 0.0077& \phantom{-} 0.2550& -0.3123&\\& \phantom{-} 0.0398& \phantom{-} 0.7264&
 -0.4035& \phantom{-} 0.1319& -0.4025&\\& \phantom{-} 0.5188& \phantom{-} 0.6385& -0.2542& \phantom{-}
  0.1980& -0.4949 & \phantom{-}   \bigl{]}
\end{matrix}$

\noindent $  \begin{matrix}
\phi_{IJ}  \phantom{ab} : \bigl{[}& -0.3303& -0.0644& \phantom{-} 0.1094& -0.2688&
 -0.2240&\\& \phantom{-} 0.3459& -0.3983& \phantom{-}  0.2833& \phantom{-} 0.1312& \phantom{-} 0.1544&\\& \phantom{-}
  0.0052& \phantom{-} 0.0824& -0.0110& -0.0018& -0.0333&\\& \phantom{-} 0.0544&
 -0.1089& \phantom{-} 0.1617& \phantom{-} 0.0938& \phantom{-} 0.1104 & \phantom{-}   \bigl{]}
\end{matrix}$

\noindent $   	\begin{matrix}
\phi_{\alpha \beta} \phantom{ab} : \bigl{[}&-0.0462& \phantom{-} 0.1916& \phantom{-}  \bigl{]}
\end{matrix}$
\vspace{0.1in}
\newline
\vspace{0.1in}
 \noindent\rule{15cm}{0.4pt}
\item Potential = -1.30461$g^2$ \newline
      $m^{2}/m_{0}^{2}(\psi) = 1.8625_2, 3.0451_4, 3.6226_2$ \newline
      SUSY = 0

\noindent $  	\begin{matrix}
\phi_{IJK \alpha} :  \bigl{[}& \phantom{-}0.3792& \phantom{-}  0.0065& \phantom{-} 0.0542& -0.0318& \phantom{-} 0.1693&\\& \phantom{-} 0.0901& \phantom{-}
  0.1697& \phantom{-} 0.5486& \phantom{-} 0.6703& -0.8252&\\& \phantom{-}0.2408& -0.3291& \phantom{-}
  0.1655& -0.2277& \phantom{-} 0.2655&\\& -1.3261& \phantom{-} 0.4648& \phantom{-} 0.4188&
 -0.5847& \phantom{-}  0.5837 & \phantom{-}   \bigl{]}
\end{matrix}$

\noindent $  \begin{matrix}
\phi_{IJ}  \phantom{ab} : \bigl{[}& \phantom{-}0.0745& \phantom{-} 0.0876& \phantom{-}  0.0608& \phantom{-} 0.1479& \phantom{-}
  0.2749&\\& -0.0973& \phantom{-}  0.0444& \phantom{-} 0.2022& -0.2839& \phantom{-} 0.0155&\\& \phantom{-}
  0.2171& -0.0549& \phantom{-} 0.2786& -0.3587& -0.2702&\\& -0.0942&
 -0.1234& \phantom{-} 0.3827& -0.3123& \phantom{-} 0.0865& \phantom{-}  \bigl{]}
\end{matrix}$

\noindent $  \begin{matrix}
\phi_{\alpha \beta} \phantom{ab} : \bigl{[}&-0.1954& -0.0395& \phantom{-}  \bigl{]} 
\end{matrix}$
\vspace{0.1in}
\newline
\vspace{0.1in}
 \noindent\rule{15cm}{0.4pt}
\item Potential = -1.31918$g^2$ \newline
      $m^{2}/m_{0}^{2}(\psi) = 2.0067_2, 2.3137_2, 3.3898_2, 3.4794_2$ \newline
      SUSY = 0

\noindent $  \begin{matrix}
\phi_{IJK \alpha} :  \bigl{[}&-0.3352& -0.1511& -0.2291& -0.4901&  -0.4108&\\& -0.5881& \phantom{-}
  0.8963& -0.1645& \phantom{-} 0.0909& \phantom{-} 0.1354&\\& -0.3969& \phantom{-} 0.3832& \phantom{-}
  0.9423& \phantom{-} 0.2324& -0.7280&\\& \phantom{-} 0.0384& \phantom{-} 0.8461& -0.5836&
 -0.1182& \phantom{-} 0.2538 & \phantom{-}   \bigl{]}
\end{matrix}$ \newline
$\begin{matrix}
\phi_{IJ}  \phantom{ab} : \bigl{[}& -0.0860& \phantom{-} 0.0510& -0.1061& -0.0345& \phantom{-}
  0.0563&\\& -0.0414& -0.1125& -0.1850& \phantom{-} 0.3208& -0.0359&\\&
 -0.0851& -0.2724& \phantom{-} 0.2732& \phantom{-} 0.5183& \phantom{-} 0.0027&\\& -0.4495&\phantom{-}
  0.2080& \phantom{-} 0.1522& -0.3976& -0.1481& \phantom{-}\bigl{]}
\end{matrix}$

\noindent $  \begin{matrix}
\phi_{\alpha \beta} \phantom{ab} : \bigl{[}&-0.1394& \phantom{-} 0.1717& \phantom{-} \bigl{]} 
\end{matrix}$
\vspace{0.1in}
\newline
\vspace{0.1in}
 \noindent\rule{15cm}{0.4pt}
\item Potential = -1.38225$g^2$ \newline
      $m^{2}/m_{0}^{2}(\psi) = 2.1565_2, 2.6257_2, 3.8238_2, 4.4247_2$ \newline
      SUSY = 0

\noindent $  \begin{matrix}
\phi_{IJK \alpha} :  \bigl{[}&-1.1020& -0.3595& -0.4405& \phantom{-} 0.5419& -0.0501&\\& \phantom{-} 0.0106&
 -0.2083& -0.5048& -0.5409& \phantom{-} 0.3079&\\& -0.0524& \phantom{-} 0.7383&
 -0.0384& \phantom{-} 0.3684& \phantom{-} 0.5508&\\& \phantom{-} 0.1843& -0.9437& \phantom{-} 0.3812&\phantom{-}
  0.1504& \phantom{-} 0.7336 & \phantom{-}  \bigl{]}
\end{matrix}$

\noindent $  \begin{matrix}
\phi_{IJ}  \phantom{ab} : \bigl{[}& \phantom{-}0.3841& -0.2175& -0.1069& -0.1276& \phantom{-}
  0.1064&\\& -0.0923& \phantom{-} 0.0651& -0.4053& \phantom{-} 0.1171& \phantom{-} 0.3588&\\&\phantom{-}
  0.2771& -0.0583& \phantom{-}  0.1246& -0.1994& -0.2135&\\& \phantom{-} 0.2667&\phantom{-}
  0.0712& \phantom{-} 0.1363& \phantom{-} 0.3093& -0.3725 & \phantom{-}\bigl{]}
\end{matrix}$

\noindent $  	\begin{matrix}
\phi_{\alpha \beta} \phantom{ab} : \bigl{[}&-0.0631&-0.0313& \phantom{-}  \bigl{]}
\end{matrix}$
\vspace{0.1in}
\newline
\vspace{0.1in}
 \noindent\rule{15cm}{0.4pt}
\item Potential = -1.39104$g^2$ \newline
      $m^{2}/m_{0}^{2}(\psi) = 2.5755_4, 3.7257_4$ \newline
      SUSY = 0

\noindent $  \begin{matrix}
\phi_{IJK \alpha} :  \bigl{[}& \phantom{-}0.5045& -0.1064& -1.1109& \phantom{-}  0.2760& \phantom{-} 0.5296&\\& \phantom{-}    0.1087&
 -0.2651& \phantom{-}0.0783& -0.3866& -0.5842&\\& \phantom{-} 0.7156& \phantom{-} 0.1431& \phantom{-}
  0.4666& -0.0244& \phantom{-} 1.1359&\\& -0.0142& -0.0181& \phantom{-} 0.0560&
 -0.7262& \phantom{-}   0.2977& \phantom{-}  \bigl{]}
\end{matrix}$

\noindent $  \begin{matrix}
\phi_{IJ}  \phantom{ab} : \bigl{[}& \phantom{-}0.0878& \phantom{-} 0.1556& -0.1148& \phantom{-} 0.4289& \phantom{-}
  0.0165&\\& -0.2806& -0.2796& \phantom{-} 0.0032& -0.3064& -0.2732&\\&
 -0.4275& -0.0099& \phantom{-} 0.2155& \phantom{-} 0.0093& -0.0511&\\& -0.0950&\phantom{-}
  0.2161& \phantom{-} 0.3355& \phantom{-} 0.0560& -0.2377 & \phantom{-}\bigl{]}
\end{matrix}$

\noindent $  \begin{matrix}
\phi_{\alpha \beta} \phantom{ab} : \bigl{[}& \phantom{-}0.0064& \phantom{-}   0.0866& \phantom{-} \bigl{]} 
\end{matrix}$
\vspace{0.1in}
\newline
\vspace{0.1in}
 \noindent\rule{15cm}{0.4pt}
\item Potential = -1.41675$g^2$ \newline
      $m^{2}/m_{0}^{2}(\psi) = 2.5411_4, 4.6817_4$ \newline
      SUSY = 0

\noindent $  \begin{matrix}
\phi_{IJK \alpha} :  \bigl{[}& \phantom{-}0.4578& \phantom{-} 0.2300& -0.5013& -0.6360& \phantom{-} 0.1031&\\& \phantom{-} 1.0931&
 -0.2719& -0.0954& -0.1685& -0.0306&\\& -0.3856&  -0.3697& \phantom{-}
  0.8386& -1.0437& -0.5426&\\& -0.5547& -0.0737& -0.0111&
 -0.6834& -0.5423 & \phantom{-}   \bigl{]}
\end{matrix}$

\noindent $  \begin{matrix}
\phi_{IJ}  \phantom{ab} : \bigl{[}&-0.1527& -0.0826& \phantom{-} 0.1660& -0.5517& \phantom{-}
  0.3393&\\& -0.2103& -0.0944& \phantom{-} 0.1184& \phantom{-} 0.0935& \phantom{-} 0.0729&\\&
 -0.0363& \phantom{-} 0.0453& \phantom{-} 0.4574& \phantom{-} 0.3743& \phantom{-} 0.1548&\\& -0.1557&
 -0.2750& -0.1443& -0.1357& \phantom{-}  0.1144 & \phantom{-}\bigl{]}
\end{matrix}$

\noindent $  \begin{matrix}
\phi_{\alpha \beta} \phantom{ab} : \bigl{[}&-0.0944& -0.1356& \phantom{-}  \bigl{]}
\end{matrix}$
\vspace{0.1in}
\newline
\vspace{0.1in}
 \noindent\rule{15cm}{0.4pt}
\item Potential = -1.41741$g^2$ \newline
      $m^{2}/m_{0}^{2}(\psi) = 2.6666_4, 4.4444_4$ \newline
      SUSY = 0

\noindent $  	\begin{matrix}
\phi_{IJK \alpha} :  \bigl{[}& \phantom{-}0.5070& \phantom{-} 0.0319& -0.1032& \phantom{-} 0.0354&  -0.3116&\\&  -0.5175& \phantom{-}
  0.9364& \phantom{-} 0.5686& \phantom{-} 0.3663& -0.2258&\\& -0.3893& \phantom{-} 0.2342&
 -0.7448& \phantom{-}  0.2160& \phantom{-} 0.7288&\\& -0.3089& -0.2455& \phantom{-} 0.9987&
 -0.4311& -0.9237 & \phantom{-}   \bigl{]}
\end{matrix}$

\noindent $  \begin{matrix}
\phi_{IJ}  \phantom{ab} : \bigl{[}& \phantom{-}0.0976& -0.1470& \phantom{-} 0.0492& -0.0833& \phantom{-}
  0.2991&\\& \phantom{-} 0.0705& -0.4666& \phantom{-} 0.4521& -0.2492& -0.0404&\\&
 -0.3528& -0.0945& -0.2121& \phantom{-} 0.2326& -0.0535&\\& -0.1107&
 -0.0161& -0.0079& -0.3507& \phantom{-}  0.0202 & \phantom{-}\bigl{]}
\end{matrix}$

\noindent $  \begin{matrix}
\phi_{\alpha \beta} \phantom{ab} : \bigl{[}& \phantom{-}0.3537& \phantom{-} 0.1304& \phantom{-}  \bigl{]}
\end{matrix}$
\vspace{0.1in}
\newline
\vspace{0.1in}
 \noindent\rule{15cm}{0.4pt}
\item Potential = -1.46065$g^2$ \newline
      $m^{2}/m_{0}^{2}(\psi) = 2.5486_2, 2.8721_2, 3.5484_2, 4.7271_2$ \newline
      SUSY = 0

\noindent $  	\begin{matrix}
\phi_{IJK \alpha} :  \bigl{[}& \phantom{-}0.8141& \phantom{-} 0.1263& -0.9595& \phantom{-} 0.0813& \phantom{-} 0.3003&\\& -0.3246& \phantom{-}
  0.3777& \phantom{-} 0.2840& \phantom{-} 0.2859& \phantom{-} 0.8808&\\& \phantom{-} 0.6213& \phantom{-} 0.3636&
 -0.7422& \phantom{-} 0.5072& -0.4096&\\& -0.2925& -0.0826& -0.3584&
 -1.0298& \phantom{-} 0.4096 & \phantom{-}   \bigl{]}
\end{matrix}$

\noindent $  \begin{matrix}
\phi_{IJ}  \phantom{ab} : \bigl{[}& \phantom{-}0.1737& \phantom{-} 0.0877& -0.2417& -0.0978& \phantom{-}
  0.1136&\\& \phantom{-} 0.4825& \phantom{-} 0.1540& \phantom{-} 0.1853& \phantom{-} 0.0108& \phantom{-} 0.1840&\\&
 -0.4475& \phantom{-} 0.3142& -0.0795& -0.0246& \phantom{-}  0.3659&\\& \phantom{-} 0.0092&
 -0.0281& \phantom{-} 0.1775& \phantom{-} 0.0829& -0.0457& \phantom{-}\bigl{]}
\end{matrix}$

\noindent $  	\begin{matrix}
\phi_{\alpha \beta} \phantom{ab} : \bigl{[}& \phantom{-}0.6531& \phantom{-} 0.3501& \phantom{-}  \bigl{]} 
\end{matrix}$
\vspace{0.1in}
\newline
\vspace{0.1in}
 \noindent\rule{15cm}{0.4pt}
\item Potential = -1.46073$g^2$ \newline
      $m^{2}/m_{0}^{2}(\psi) = 2.5779_2, 2.9533_2, 3.4325_2, 4.6655_2$ \newline
      SUSY = 0

\noindent $  	\begin{matrix}
\phi_{IJK \alpha} :  \bigl{[}& \phantom{-}0.4143& -0.5478& \phantom{-} 0.9046& \phantom{-} 0.3579& -0.0142&\\& -0.5180&
 -0.3720& \phantom{-} 1.2699& \phantom{-} 0.6505& \phantom{-} 0.3487&\\& \phantom{-} 0.1898& -0.8265&
 -0.1295& -0.1092& \phantom{-} 0.1097&\\& \phantom{-} 0.5064& \phantom{-} 0.1292& \phantom{-} 0.2429&
 -0.4892& \phantom{-} 1.0179 & \phantom{-}   \bigl{]}
\end{matrix}$

\noindent $  \begin{matrix}
\phi_{IJ}  \phantom{ab} : \bigl{[}& \phantom{-}0.2617& -0.0752& \phantom{-} 0.3280& \phantom{-} 0.0231&
 -0.0322&\\& -0.1835& -0.0262&  -0.0678& \phantom{-} 0.1579& \phantom{-}  0.1063&\\& \phantom{-}
  0.0523& -0.2809& \phantom{-} 0.3321& -0.1705& \phantom{-} 0.2598&\\& \phantom{-} 0.1327& \phantom{-}
  0.1924& -0.2024& \phantom{-} 0.2580& -0.0564 & \phantom{-}\bigl{]}
\end{matrix}$

\noindent $  \begin{matrix}
\phi_{\alpha \beta} \phantom{ab} : \bigl{[}&-0.0311& \phantom{-} 0.2936& \phantom{-}  \bigl{]}
\end{matrix}$
\vspace{0.1in}
\newline
\vspace{0.1in}
 \noindent\rule{15cm}{0.4pt}
\item Potential = -1.49704$g^2$ \newline
      $m^{2}/m_{0}^{2}(\psi) = 2.5643_2, 2.7441_2, 4.9241_2, 4.9868_2$ \newline
      SUSY = 0

\noindent $  	\begin{matrix}
\phi_{IJK \alpha} :  \bigl{[}& \phantom{-}0.0109& -0.0719& -0.0815& \phantom{-} 0.5474&
 -0.2377&\\& \phantom{-} 0.2254& \phantom{-} 0.1968& -0.9845& \phantom{-}
  0.1016& -0.4452&\\& \phantom{-} 1.6962& \phantom{-} 0.4204& \phantom{-}
  0.7246& -0.3013& -0.0778&\\& -0.1267&
 -0.6545& -0.5041& -0.0784& \phantom{-} 0.7185 & \phantom{-}  \bigl{]}
\end{matrix}$

\noindent $  \begin{matrix}
\phi_{IJ}  \phantom{ab} : \bigl{[}& \phantom{-}0.0964& \phantom{-} 0.0903& \phantom{-} 0.0390& \phantom{-} 0.0055&
 -0.2337&\\& \phantom{-} 0.2988& \phantom{-} 0.0958& -0.2709& \phantom{-}
  0.1139& \phantom{-} 0.1907&\\& \phantom{-} 0.2990& \phantom{-} 0.0509& \phantom{-}
  0.0269& -0.4756& \phantom{-} 0.4467&\\& -0.0014& \phantom{-}
  0.0119& \phantom{-} 0.1248& -0.1364& \phantom{-} 0.3742 & \phantom{-}\bigl{]}
\end{matrix}$

\noindent $  \begin{matrix}
\phi_{\alpha \beta} \phantom{ab} : \bigl{[}& \phantom{-}0.0637& -0.2613& \phantom{-}  \bigl{]}
\end{matrix}$
\vspace{0.1in}
\newline
\vspace{0.1in}
 \noindent\rule{15cm}{0.4pt}
\item Potential = -1.49967$g^2$ \newline
      $m^{2}/m_{0}^{2}(\psi) = 2.7963_4, 4.6705_4$ \newline
      SUSY = 0

\noindent $  \begin{matrix}
\phi_{IJK \alpha} :  \bigl{[}&-0.5470& -0.3379& \phantom{-} 0.6857& -0.3440& \phantom{-} 0.5168&\\& -0.1625&-1.0394& -0.0909& -0.6156& -0.3561&\\& \phantom{-} 0.0076& -1.0169& \phantom{-}
  0.1301& -0.4126& -0.6633&\\& \phantom{-}  0.1030& \phantom{-} 0.3956& -0.4765& \phantom{-}
  0.3572& -0.6718 & \phantom{-}   \bigl{]}
\end{matrix}$

\noindent $  \begin{matrix}
\phi_{IJ}  \phantom{ab} : \bigl{[}& \phantom{-}0.5445& \phantom{-} 0.2800& -0.0249& -0.0059& \phantom{-}
  0.2232&\\& -0.0843& \phantom{-} 0.6194& -0.2070& \phantom{-} 0.0222& -0.2139&\\&
 -0.0774& \phantom{-} 0.3875& \phantom{-} 0.2824& \phantom{-} 0.0537& -0.2025&\\& \phantom{-} 0.5864&
 -0.2068& -0.2944& \phantom{-} 0.4541& \phantom{-} 0.2089 & \phantom{-}\bigl{]}
\end{matrix}$

\noindent $  \begin{matrix}
\phi_{\alpha \beta} \phantom{ab} : \bigl{[}& \phantom{-}0.8443& \phantom{-}   0.4465& \phantom{-}  \bigl{]}
\end{matrix}$
\vspace{0.1in}
\newline
\vspace{0.1in}
 \noindent\rule{15cm}{0.4pt}
\item Potential = -1.50186$g^2$ \newline
      $m^{2}/m_{0}^{2}(\psi) = 2.9579_4, 4.3614_4$ \newline
      SUSY = 0

\noindent $  	\begin{matrix}
\phi_{IJK \alpha} :  \bigl{[} & \phantom{-}0.6661& \phantom{-} 0.1872& -1.1302& -0.2231& -0.6673&\\& \phantom{-} 1.0202&
 -0.7304& -0.6898& \phantom{-} 0.3030& -0.0776&\\& \phantom{-} 0.7682& -0.1960& \phantom{-}
  0.2955&-0.2041& -0.5748&\\& \phantom{-}  0.0011& -0.6541& \phantom{-} 0.3931&\phantom{-}
  0.2222& \phantom{-} 0.6468 & \phantom{-}   \bigl{]}
\end{matrix}$

\noindent $  \begin{matrix}
\phi_{IJ}  \phantom{ab} : \bigl{[}&-0.1079& \phantom{-} 0.1653& -0.3424& \phantom{-} 0.0965&
 -0.2137&\\& -0.6426& \phantom{-} 0.0919& \phantom{-} 0.3443& -0.2391& -0.1141&\\& \phantom{-}
  0.1506& -0.0666& -0.0747& -0.0829& -0.2133&\\& -0.1030& \phantom{-}
  0.2672& \phantom{-} 0.0681& -0.0529& -0.0405 & \phantom{-}\bigl{]}
\end{matrix}$

\noindent $  \begin{matrix}
\phi_{\alpha \beta} \phantom{ab} : \bigl{[}&-0.0572& -0.2135& \phantom{-} \bigl{]}
\end{matrix}$
\vspace{0.1in}
\newline
\vspace{0.1in}
 \noindent\rule{15cm}{0.4pt}
\item Potential = -1.5109$g^2$ \newline
      $m^{2}/m_{0}^{2}(\psi) = 2.3407_2, 2.5155_2, 5.2521_2, 5.5349_2$ \newline
      SUSY = 0

\noindent $  \begin{matrix}
\phi_{IJK \alpha} :  \bigl{[}&-0.5647& \phantom{-} 0.1444& \phantom{-} 1.4551& -0.0401& \phantom{-} 0.5691&\\& \phantom{-} 0.3203&
 -0.2771& -0.9053& \phantom{-} 0.0201& -0.4700&\\& \phantom{-}  0.5849& \phantom{-} 0.4366&
 -0.3968&-0.2621& \phantom{-}   0.4676&\\& -1.0103& \phantom{-} 0.2259& -0.4547& \phantom{-}
  0.0375& -0.1095 & \phantom{-}  \bigl{]}
\end{matrix}$

\noindent $  \begin{matrix}
\phi_{IJ}  \phantom{ab} : \bigl{[}&-0.0738& \phantom{-} 0.0752& \phantom{-}  0.4773& -0.2479& \phantom{-}
  0.1554&\\& \phantom{-}0.4256& \phantom{-} 0.2557& -0.1201& \phantom{-} 0.0073& \phantom{-} 0.0333&\\&
 \phantom{-} 0.1572& \phantom{-}  0.0618& \phantom{-} 0.4535& -0.0586& -0.0733&\\& \phantom{-} 0.0047& \phantom{-}
  0.1273&  -0.3006& \phantom{-} 0.3402& -0.0354& \phantom{-}\bigl{]}
\end{matrix}$

\noindent $  \begin{matrix}
\phi_{\alpha \beta} \phantom{ab} : \bigl{[}&-0.0679& -0.3911& \phantom{-}  \bigl{]}
\end{matrix}$
\vspace{0.1in}
\newline
\vspace{0.1in}
 \noindent\rule{15cm}{0.4pt}
\item Potential = -1.54778$g^2$ \newline
      $m^{2}/m_{0}^{2}(\psi) = 2.9185_2, 3.5111_4, 4.6963_2$ \newline
      SUSY = 0

\noindent $   	\begin{matrix}
\phi_{IJK \alpha} : \bigl{[}& \phantom{-}0.1890& -0.5513& -1.2676& \phantom{-} 0.5408& -0.7060&\\& -0.1871&
 -0.0405& \phantom{-} 0.6394& -0.1154& \phantom{-} 0.7851&\\& -0.0294& -0.2043&
 -0.3016& \phantom{-} 0.2793& \phantom{-} 0.5621&\\& -1.0157& -0.2272& -0.1133& \phantom{-}
  0.1936& \phantom{-} 0.3877& \phantom{-}   \bigl{]}
\end{matrix}$

\noindent $  \begin{matrix}
\phi_{IJ}  \phantom{ab} : \bigl{[}& \phantom{-}0.5632& \phantom{-} 0.3586& \phantom{-}  0.0117& \phantom{-} 0.1890& \phantom{-}
  0.1783&\\& \phantom{-} 0.1757& \phantom{-} 0.3081& \phantom{-}  0.2079&  -0.1360& -0.5055&\\&
 -0.1263& \phantom{-} 0.5341& -0.0976& \phantom{-} 0.0022& \phantom{-} 0.4743&\\& \phantom{-} 0.5902& \phantom{-}
  0.0924& -0.0155& \phantom{-} 0.3127& \phantom{-} 0.2326& \phantom{-}\bigl{]}
\end{matrix}$

\noindent $   	\begin{matrix}
\phi_{\alpha \beta} \phantom{ab} : \bigl{[}& \phantom{-}0.3217& \phantom{-} 0.4736& \phantom{-} \bigl{]}
\end{matrix}$
\vspace{0.1in}
\newline
\vspace{0.1in}
 \noindent\rule{15cm}{0.4pt}
\item Potential = -1.73841$g^2$ \newline
      $m^{2}/m_{0}^{2}(\psi) = 2.3802_2, 3.1057_2, 5.2046_2, 5.3216$ \newline
      SUSY = 0

\noindent $   	\begin{matrix}
\phi_{IJK \alpha} :  \bigl{[}& \phantom{-}0.0584& \phantom{-} 0.1475& -0.6570& -0.1437& \phantom{-} 0.6920&\\&   -0.7354&
 -0.8385& -0.0914& -0.2961& \phantom{-} 0.5829&\\& \phantom{-} 0.6761& -0.5762& \phantom{-}
  1.4396& -0.6332& \phantom{-} 0.4682&\\& -1.0215& \phantom{-} 0.2002& \phantom{-} 0.8583&
 -0.2963& -0.2243 & \phantom{-}   \bigl{]}
\end{matrix}$

\noindent $  \begin{matrix}
\phi_{IJ}  \phantom{ab} : \bigl{[}&-0.0841& -0.3372& -0.3356& -0.0170&
 -0.2933&\\& \phantom{-} 0.0372& -0.2104& -0.5435& -0.0674& -0.4590&\\& \phantom{-}
  0.0163& -0.3045&  -0.1094& -0.5008& -0.0098&\\& \phantom{-}  0.0023&
 -0.0648& \phantom{-} 0.0154& -0.2007& -0.0209& \phantom{-}\bigl{]}
\end{matrix}$

\noindent $  	\begin{matrix}
\phi_{\alpha \beta} \phantom{ab} : \bigl{[}& -0.0854& -0.1851& \phantom{-} \bigl{]}
\end{matrix}$
 \end{enumerate}

\end{document}